\documentclass[11pt, oneside, a4paper]{article}

\usepackage[left=2cm,right=2cm,top=2cm,bottom=2cm]{geometry}
\usepackage{amsmath}
\usepackage{amsfonts}

\usepackage{mathtools}
\usepackage{bm}
\usepackage{setspace}
\usepackage{float}
\usepackage{bbm}
\usepackage{tikz}
\usepackage{enumitem}

\usepackage{authblk}   

\usepackage{url}

\usepackage[title]{appendix}

\usepackage{array}
\newcolumntype{C}[1]{>{\centering\let\newline\\\arraybackslash\hspace{0pt}}m{#1}}

\newcommand\Tstrut{\rule{0pt}{2.6ex}}         
\newcommand\Bstrut{\rule[-0.9ex]{0pt}{0pt}}   

\allowdisplaybreaks

\title{\textbf{Online control of the false discovery rate in biomedical research}}

\author[1]{D.\ S.\ Robertson}
\author[1,2]{J.\ M.\ S.\ Wason}
\affil[1]{\small MRC Biostatistics Unit, University of Cambridge, Cambridge, UK}
\affil[2]{\small Institute of Health and Society, Newcastle University, Newcastle, UK}

\date{}

\begin{document}

\maketitle

\begin{abstract}


Modern biomedical research frequently involves testing multiple related hypotheses, while maintaining control over a suitable error rate. In many applications the false discovery rate (FDR), which is the expected proportion of false positives among the rejected hypotheses, has become the standard error criterion. Procedures that control the FDR, such as the well-known Benjamini-Hochberg procedure, assume that all $p$-values are available to be tested at a single time point. However, this ignores the sequential nature of many biomedical experiments, where a sequence of hypotheses is tested without having access to future $p$-values or even the number of hypotheses. Recently, the first procedures that control the FDR in this \textit{online} manner have been proposed by Javanmard and Montanari (\textit{Ann. Stat.} 2018), and built upon by Ramdas et al.~(\textit{NIPS} 2017, \textit{ICML} 2018). In this paper, we compare and contrast these proposed procedures, with a particular focus on the setting where the $p$-values are dependent. We also propose a simple modification of the procedures for when there is an upper bound on the number of hypotheses to be tested. Using comprehensive simulation scenarios and case studies, we provide recommendations for which procedures to use in practice for online FDR control.

\end{abstract}


\vspace{3cm}

\noindent Address correspondence to D.\ S.\ Robertson,  MRC Biostatistics Unit, University of Cambridge, IPH Forvie Site, Robinson Way, Cambridge CB2 0SR, UK; E-mail: david.robertson@mrc-bsu.cam.ac.uk

\newpage


\section{Introduction}
\label{sec:intro}








In modern biomedical experiments, it is now commonplace to make a large number of related data-based decisions, formalised through statistical hypothesis testing. For example, in genomics it is routine to test hundreds of thousands of genetic variants for an association with a particularly phenotypic trait. However, performing a large number of hypothesis tests naturally gives rise to the problem of multiple comparisons~\cite{Tukey1953}: given a collection of multiple hypotheses to be tested, the goal is to distinguish which hypotheses are null and non-null, \textit{while controlling a suitable error rate}. This error rate is generally formed around the probability of incorrectly classifying a null hypothesis as non-null. Typically, a $p$-value is calculated for each hypothesis and is then used to decide whether to reject the hypothesis (and effectively declare it non-null). Multiple hypothesis testing is one of the core problems in statistical inference, and has led to a wide range of procedures that can be used to `correct' for multiplicity and ensure that a suitable error rate is controlled. In contrast, uncorrected hypothesis testing contributes to serious concerns over reproducibility, publication bias and `p-hacking' in scientific research~\cite{Ioannidis2005, Head2015}.




In confirmatory clinical contexts, where the aim is to provide definitive results, the \textit{familywise error rate} (FWER) is the typical error criterion. This ensures that the maximum probability of falsely rejecting any of the hypotheses is controlled. However, the FWER is a stringent error rate to control, and can lead to testing procedures with a substantially reduced power compared with uncorrected testing, particularly when there is a large number of hypotheses. As well, the FWER is most appropriate when the conclusion of an experiment is likely to be erroneous if at least one of the individual rejections is erroneous (e.g.\ when taking forward a single treatment in a clinical trial). This may not be the case in many trial designs. For example, in a screening study multiple genes may be truly associated with a disease and the aim is to find as many of these genes as possible, without an overall decision having to be made (with the candidate genes discoveries being validated in a follow-up replication study).


To address these concerns about the FWER, the \textit{false discovery rate} (FDR) was introduced by~\cite{Benjamini1995}. The FDR is the expected \textit{proportion} of the discoveries (i.e.\ rejections) that are false.  Hence it `scales' as the number of hypothesis tested increases. The FDR is now the error criterion of choice for large-scale multiple hypothesis testing, particularly in genomics and brain imaging. With the increase in the multiplicity and complexity of the objectives and structure of modern clinical trials, there is much scope to apply FDR in this context too.




Traditionally, multiple hypothesis testing is \textit{offline} in nature, in the sense that a procedure for testing~$N$ hypotheses will receive all of the $p$-values $(p_1, \ldots, p_N)$ at once. Step-up and step-down multiple testing procedures (for example) order the $p$-values by size, and so \textit{require} all $p$-values in order to work. However, in many areas of biomedical research this offline paradigm is not appropriate, as hypotheses are tested in a sequential manner. Two key examples (which are expanded upon in Section~\ref{sec:motivation}) are public biological databases and perpetual platform trials. What is needed therefore are procedures for \textit{online} hypothesis testing. More precisely, the online hypothesis testing problem is defined as follows~\cite[pg.~2]{Javanmard2015}:\\[-16pt]

\textit{Hypotheses arrive sequentially in a stream. At each step, the analyst must decide whether to reject the current null hypothesis without having access to the number of hypotheses (potentially infinite) or the future p-values, but solely based on the previous decisions.} \\[-16pt]


Until recently, it was not known how to provably control the FDR for online hypothesis testing. The first work that guaranteed online FDR control was by Javanmard and Montanari~\cite{Javanmard2015, Javanmard2018}. They looked at a class of generalised alpha investing procedures proposed in~\cite{Aharoni2014}, and proved that any rule in this class controls the FDR in the online setting, provided that the $p$-values for the true null hypotheses are \textit{independent} from each other. They also proposed procedures that provably control the online FDR under general $p$-value dependencies. Their work was the basis for further recently proposed procedures for online FDR control by Ramdas et al.~\cite{Ramdas2017, Ramdas2018}, again in the setting with independent $p$-values.


The aim of this paper is to comprehensively compare and contrast the different proposed procedures in~\cite{Javanmard2015, Javanmard2018, Ramdas2017, Ramdas2018}, with a particular focus on the performance (in terms of the FDR and power of the procedures) under \textit{dependent} $p$-values. This focus is because in practice, it is not appropriate to assume independence between the $p$-values in many biomedical applications. For example, gene expression data may be correlated due to linkage disequilibrium, while in the clinical trial setting, the estimates of treatment differences will be correlated by definition when comparing with a common control.  Hence, it is important to explore how robust the procedures are under violations of independence, as well as to see how they compare with the procedures designed for general dependencies.

Another area we focus on is how the procedures perform when the number of hypotheses~$N$ is relatively small (i.e.\ $N < 1000$), since previous work has assumed $N \geq 1000$.  A relatively small~$N$ would be a more relevant setting for clinical trials, for example. We also propose a simple modification of the procedures for when it is possible to identify an upper bound for the number of hypotheses to be tested, which can result in a substantial increase in the power of the procedure.


The structure of the rest of the paper is as follows. Section~\ref{sec:motivation} describes two motivating examples where online hypothesis testing would naturally be encountered. Section~\ref{sec:error_rates} presents formal definitions of error rate control, and Section~\ref{sec:online_control} describes the procedures for online FDR control in detail. Section~\ref{sec:simul} presents a comprehensive simulation study of the procedures under dependent $p$-values, while Section~\ref{sec:case_study} presents two case studies of applying online FDR control. We give our recommendations and conclusions in Section~\ref{sec:discuss}.


\section{Motivating examples}
\label{sec:motivation}

Online hypothesis testing is a naturally occurring phenomenon in biomedical research, and we present two key examples below.

\subsection{Public biological databases}
\label{subsec:public_database}

Public databases and shared data resources are becoming increasingly pervasive and important in modern biomedical research, particularly in the fields of genetics and molecular biology. Some well-known examples include the 1000 genomes project~\cite{1000genomes2015}, which resulted in a large public catalogue of human variation and genotype data; the Wellcome Trust Case Control Consortium~\cite{WTCC2007}, which collects large-scale data for whole-genome association studies and provides them to research groups upon request; and ArrayExpress~\cite{Kolesnikov2014}, which is a major repository that archives functional genomics data from microarray and sequencing platforms. Another example is the International Mouse Phenotyping Consortium database~\cite{Koscielny2013, Dickinson2016}, which we describe as one of our case studies in Section~\ref{sec:case_study}. 

Multiple testing naturally occurs in this setting in two ways. Firstly, such databases can be accessed by multiple independent researchers at different times. When a researcher or research group comes up with a new hypothesis, they can simply fetch the relevant data from a database and perform a statistical test.
Secondly, in many databases the family of hypotheses to be tested grows over time as new data is continually added. In both of these scenarios, the number of hypotheses being tested will be unknown and potentially very large. 

In order to control the number or proportion of false discoveries in this context, a procedure needs to be sequential, allowing a researcher to decide whether to reject a current hypothesis with minimal information about previous hypotheses, and without prior knowledge of even the number of hypotheses that are going to be tested in the future. This is precisely the online hypothesis testing framework described in Section~\ref{sec:intro}.








\subsection{Perpetual platform trials}

\begin{figure}[ht!]

\begin{tikzpicture}

\begin{scope}[thick]

	\draw[] (-6,7) node[left, black] {T1} -- (-1,7) node[right]{$H_1$};     

	\draw[] (-6,6.2) node[left, black] {T2} -- (-2,6.2) node[right]{$H_2$};         
                                        
    \draw[] (-6,5.4) node[left, black] {T3} -- (1.4,5.4) node[right]{$H_3$};     
    
    \draw[] (-3.5,4.6) node[left, black] {T4} -- (4,4.6) node[right]{$H_5$};         
                                        
    \draw[] (-1.5,3.8) node[left] {T5} -- (2.5,3.8) node[right]{$H_4$};         
                                        
	\draw[] (0.2,3) node[left] {T6} -- (6.3,3) node[right]{$H_6$};    
                                        
	\draw[] (1.8,2.2) node[left] {T7} -- (9,2.2) node[right] {$\cdots$};                            

	\draw[] (2.4,1.4) node[left, black] {T8} -- (9,1.4) node[right] {$\cdots$};                                                                             
                                                                 
    \foreach \x in {4,6,8}{
	\node[] at (\x,0.6) {$\vdots$};
	}                                      
	
    \draw[] (-6,-0.2) node[left] {C1} -- (4,-0.2);
                                        
	\draw[] (3.8,-1) node[left] {C2} -- (9,-1) node[right] {$\cdots$};   
                                                                               
    \foreach \x in {4,6,8}{
	\node[] at (\x,-1.8) {$\vdots$};
	
	\draw[->, line width = 2] (-6,-3) -- (10,-3);
	\node[] at (1.5, -3.5) {\textbf{Time}};
		
	}
                               
\end{scope}
\end{tikzpicture}

\caption{\label{platform_trial} Diagram of a perpetual platform trial, where T = experimental treatment, C = control and $H_i$ represents the $i$-th hypothesis to be tested.
\label{fig:platform_trial}} 
\end{figure}
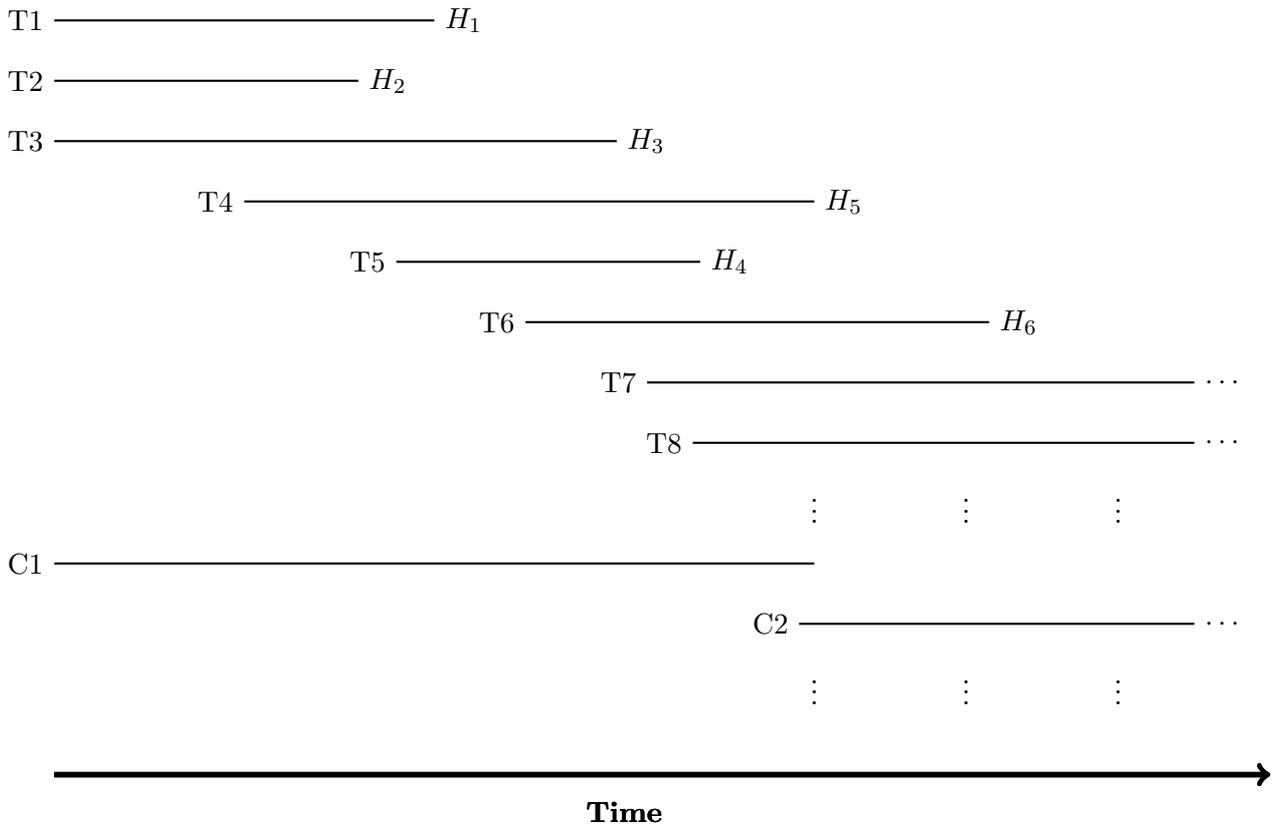

A perpetual platform trial has a single master protocol that evaluates multiple treatments across one or more patient types, and allows a potentially large number of treatments to be added during the course of the trial~\cite{Saville2016}. Treatments are dropped from the trial after they have been formally tested for effectiveness. Such a trial is (in theory) `perpetual' in that new treatments can always continue to enter into the trial and be tested. Figure~\ref{fig:platform_trial} gives a diagrammatic representation of a typical perpetual platform trial. \vspace{6pt}

In a platform trial, treatments are introduced at different time points by design. However, the trial investigators will wish to make a decision on whether a treatment is beneficial as soon as the data are ready. In particular, decisions are made without waiting for results from the other treatment arms.
%
Hence, the treatment effects are tested sequentially in an online manner, where the number of treatments to be tested in the future is unknown. More formally, a platform trial generates a sequence of null hypotheses $(H_1, H_2, H_3, \ldots)$ which are tested sequentially. Hypothesis $H_i$ tests the value of some parameter~$\theta_i$, such as an estimate of the treatment difference compared to a control arm. Note that the control arm could also change during the trial (like in Figure~\ref{fig:platform_trial}), for example due to one of the previously evaluated experimental arms being confirmed as effective and becoming the new control.

Rejecting~$H_i$ would mean stopping treatment arm~$i$ for efficacy and `graduating' the treatment to a phase~III confirmatory trial. Conversely, accepting (i.e.\ failing to reject) $H_i$ would mean stopping treatment arm~$i$ for futility. 
%
It is important to note that the $p$-values generated from the platform trial described above will \textit{not} in general be independent. Dependencies will primarily arise due to the shared control data that is re-used to test multiple hypotheses. 


A current example of a long-running platform trial is the STAMPEDE trial~\cite{Parmar2017, Sydes2018} for patients with locally advanced or metastatic prostate cancer. In 2005, the trial started with five comparisons against the standard-of-care, the results of which have now all been published. Since then, three arms have been added which are now closed to recruitment, with an additional two arms currently under evaluation. Conceivably, in the future further arms will be added to the trial as new treatments become available.



\section{Error rates}
\label{sec:error_rates}

In this section, we formally define the FWER and FDR, and introduce the Benjamini-Hochberg procedure for controlling the FDR. Let $\mathcal{H}(n) = (H_1, \ldots, H_n)$ denote the first~$n$ hypotheses tested. 
A general testing procedure provides a sequence of test levels~$\alpha_i$ with decision rule \[
R_i = \begin{cases}
1 \qquad \text{if } p_i \leq \alpha_i \qquad \text{(reject } H_i) \\
0 \qquad \text{otherwise} \qquad \text{(accept } H_i)
\end{cases}
\]
Also let $R(n) = \sum_{i=1}^n R_i$ denote the number of rejected hypotheses in~$\mathcal{H}(n)$, and $V^{\theta}(n)$ denote the number of \textit{incorrectly} rejected hypotheses in~$\mathcal{H}(n)$. 

The FWER is the probability of falsely rejecting any null hypothesis from~$\mathcal{H}(n)$:
\[
\text{FWER}(n) \equiv  \mathbb{P} \left( V(n) \geq 1 \right) 
\]


The FDR is the expected false discovery proportion (FDP), i.e.\ the expected proportion of false discoveries (incorrectly rejected hypotheses) among the rejected hypotheses in~$\mathcal{H}(n)$:
\[
\text{FDR}(n) \equiv \mathbb{E}(\text{FDP}(n)) \equiv \mathbb{E} \left( \frac{V(n)}{\max(R(n), 1)} \right)
\]

The FWER can be controlled at level~$\alpha$ in a simple manner by using a Bonferonni-type correction. More precisely, we can choose different significance levels~$\alpha_i$ for~$H_i$, such that $\sum_{i=1}^{\infty} \alpha_i = \alpha$. Examples include setting $\alpha_i = \alpha / 2^{i}$ or $\alpha_i = 6\alpha/i^2\pi^2$. Note that this procedure only requires knowledge of the number of tests performed before the current one. However, this method suffers from a low statistical power, with the probability of the null hypothesis~$H_n$ being rejected tending to zero as $n$ increases.

Benjamini and Hochberg developed the first procedure to control the FDR, assuming that all $p$-values are known. Given $p$-values $p_1, p_2, \ldots, p_N$ and a significance level~$\alpha$, the Benjamini-Hochberg (BH) procedure is as follows~\cite{Benjamini1995}:
\begin{enumerate}
\item Order the $p$-values from smallest to largest, giving ordered $p$-values $p_{(1)} \leq \ldots \leq p_{(N)}$, and define $p_{(0)} = 0$
\item Let $i^*$ be the maximal index such that $p_{(i^*)} \leq i \alpha / N $
\item Reject $H_j$ for every test with $p_j \leq p_{(i^*)}$
\end{enumerate}

In contrast to the Bonferonni-type procedure for controlling the FWER, the BH procedure requires knowledge of all the $p$-values. Hence, it does not address the situation where we do not know the number of hypotheses or future $p$-values, such as in the perpetual platform trial context.
Instead, what we require is online control of the FDR, where the test levels can only be functions of prior decisions: $\alpha_i = \alpha_i(R_1, R_2, \ldots, R_{i-1})$.


\section{Online rules for controlling the false discovery rate}
\label{sec:online_control}

We now describe the proposed procedures for online control of the FDR, for both independent and dependent test statistics. We also propose a simple modification to the procedures for when there is an upper bound on the number of hypotheses to be tested. Some of these methods have recently been implemented in the R~package~\textit{onlineFDR}~\cite{Robertson2018}, which is is freely available through Bioconductor at \url{http://www.bioconductor.org/packages/onlineFDR}.

\subsection{LORD}

LORD stands for Levels based On Recent Discovery, and was first conceptualised in the 2015 paper by Javanmard and Montanari~\cite{Javanmard2015}. The LORD procedures are instances of generalised alpha-investing (GAI) rules proposed by Aharoni and Rosset~\cite{Aharoni2014}, and hence have an intuitive interpretation. Any GAI rule begins with an error budget, or \textit{alpha-wealth}, that is allocated to the different hypothesis tests over time. That is, there is a price (or penalty) to be paid each time a hypothesis is tested. However, when a new discovery is made (i.e.\ a hypothesis is rejected) some of the alpha-wealth is earned back, which can be viewed as a return on the alpha-investment. Hence as long as discoveries continue to be made, hypotheses can be tested indefinitely without the test levels tending towards zero.

We now introduce some notation for what follows. Let $\tau_k$ denote the time of the $k$-th  rejection: \[
\tau_k = \min \left\{ j \in \mathbb{N} : \sum_{i = 1}^j R_i = k\right\}
\] where we set $\tau_0 = 0$. Also let $\tilde{\tau}_i$ denote the last time a discovery was made before time~$i$ (where time~$i$ refers to the $i$-th hypothesis being tested): \[
\tilde{\tau}_i = \max \left\{ l \in \{1, \ldots, i - 1\} : R_l = 1\right\}
\] where we set $\tilde{\tau}_1 = 0$, and if $R_l = 0$ for $l \in \{1, \ldots, i-1\}$ then $\tilde{\tau_i} = 0$ also.


\subsubsection{Independent test statistics}
\label{subsubsec:indep}

Javanmard and Montanari~\cite{Javanmard2018} presented three versions of LORD for independent $p$-values (or equivalently, independent test statistics). Below we describe the two versions that have the highest power, called LORD~2 and LORD~3.  We first choose a non-increasing sequence of non-negative numbers $\bm{\gamma} = (\gamma_i)_{i \in \mathbb{N}}$ such that $\sum_{i=1}^{\infty} \gamma_i = 1$. In Section~\ref{subsec:seq_examples} we give some concrete examples of these sequences. We also choose two constants $w_0 \geq 0$ and $b_0 > 0$, such that $w_0 + b_0 \leq \alpha$, where more details on their interpretation are given below. 

%

\paragraph*{LORD 2}
\hspace{6pt} The test levels are based on all previous discovery times, with
\begin{equation}
\alpha_i = \gamma_i w_0 + b_0 \sum_{j \, : \, \tau_j < i} \gamma_{i - \tau_j}
\label{eq:LORD2}
\end{equation} 

Hence, up until the first discovery time, the test levels are a fraction $\gamma_i$ of~$w_0$. After each discovery time $\tau_j$, all subsequent test levels are increased by a fraction $\gamma_{i - \tau_j}$ of~$b_0$.

\paragraph*{LORD 3}
\hspace{6pt} The test levels depend on the number of hypotheses tested since the last discovery time, and the alpha-wealth accumulated at that time, with
\[
\alpha_i  = \gamma_{i - \tilde{\tau}_i} W(\tilde{\tau}_i)
\]
Here $\{W(j)\}_{j \geq 0}$ represents the alpha-wealth available at time $j$, and is defined recursively:
\begin{align}
W(0) & = w_0 \nonumber \\
W(j) & = W(j-1) - \alpha_j + b_0 R_j
\label{eq:wealth_update}
\end{align}
where $w_0$ represents the initial alpha-wealth and $b_0$ represents the amount earned for rejecting a hypothesis. 

Javanmard and Montanari proved that LORD~2 controls the FDR at level~$\alpha$ for independent $p$-values provided that $w_0 + b_0 \leq \alpha$, where $w_0 \geq 0$ and $b_0 > 0$. They also showed empirically (i.e.\ through simulation studies) that LORD~3 controls the FDR. 

\paragraph*{LORD++} 
\hspace{6pt} Ramdas et al.~\cite{Ramdas2017} derived a modified version of LORD~2, called LORD++, which has a uniformly higher power than the LORD~2 procedure. In LORD++, the test levels are chosen as follows:
\begin{equation}
\alpha_i = \gamma_i w_0 + (\alpha - w_0) \gamma_{i - \tau_1} + \alpha \sum_{j \, : \, \tau_j < i, \, \tau_j \neq \tau_1} \gamma_{i - \tau_j}
\label{eq:LORD++}
\end{equation}
 
\noindent The test levels for LORD++  will always be equal to or greater than those for LORD~2. This can be seen by simply comparing equations~\eqref{eq:LORD2} and~\eqref{eq:LORD++} for LORD~2 and LORD++ respectively, and noting that  $b_0 \leq \alpha - w_0 \leq \alpha$.

\paragraph*{SAFFRON} 

\hspace{6pt}  Finally, Ramdas et al.~\cite{Ramdas2018} derived an adaptive version of LORD++ called SAFFRON (Serial estimate of the Alpha Fraction that is Futilely Rationed On true Null hypotheses), which is based on an estimate of the proportion of true null hypotheses. More precisely, SAFFRON sets the test levels based on an estimate of the amount of alpha-wealth that is allocated to testing the hypotheses that are truly null.

To this end, we specify a constant $\lambda \in (0,1)$ and define the candidate $p$-values as those that satisfy $p_j \leq \lambda$. These candidates can be thought of as the hypotheses that are a-priori more likely to be non-null. The SAFFRON procedure then proceeds as follows:
\begin{enumerate}

\item Set the initial alpha-wealth $w_0 < (1 - \lambda) \alpha$

\item At each time~$i$, define the number of candidates after the $k$-th rejection as \[C_{k+} = C_{k+}(i) = \sum_{j = \tau_k + 1}^{i-1} C_j\] where $C_j= \mathbbm{1}\{ p_j \leq \lambda\}$ is the indicator for candidacy.

\item SAFFRON starts with $\alpha_1 = \min\{ \gamma_1 w_0, \lambda \}$. Subsequent test levels are chosen as $\alpha_i = \min \{ \lambda, \tilde{\alpha_i} \}$, where

\end{enumerate}

 \begin{equation}
\tilde{\alpha}_i = w_0 \gamma_{i - C_{0+}} + \left( (1 - \lambda) \alpha - w_0 \right) \gamma_{i - \tau_1 - C_{1+}} + (1 - \lambda)\alpha \sum_{j \, : \, \tau_j < i, \, \tau_j \neq \tau_1} \gamma_{i - \tau_j - C_{j+}}
\label{eq:SAFFRON}
\end{equation}

Hence, SAFFRON starts off with alpha-wealth $w_0$ and does not lose any of this wealth when testing candidate $p$-values~\cite{Ramdas2018}. It gains an alpha-wealth of  $((1-\lambda)\alpha - w_0)$ at the first discovery, and then $(1-\lambda)\alpha$ for each subsequent discovery. SAFFRON can make more rejections than LORD++ if there is a significant fraction of non-nulls and the signals are strong.

\paragraph{Upper bound on the number of hypotheses}
\hspace{6pt} For all the LORD methods described above, if there is an upper bound $N$ on the number of hypotheses tested, then we can simply choose $\bm{\gamma} = (\gamma_i)_{i \in \{1, \ldots, N \}}$  such that $\sum_{i = 1}^N  \gamma_i = 1$. A potential issue is that the upper bound~$N$ may need to be changed to $N'$ (say) as the trial progresses. For example, the experimenter may want to test more hypotheses than originally planned.  Suppose the change to  an upper bound of $N'$ happens immediately after $n< N$ hypotheses have been tested. The change can be accommodated in a straightforward manner by choosing new coefficients  $\bm{\gamma'} = (\gamma_i')_{i \in \{n+1, \ldots, N' \}}$ such that $\sum_{i=n+1}^{N'} \gamma_i'  = 1 - \sum_{i=1}^{n} \gamma_i$.


\subsubsection{Dependent test statistics}

Javanmard and Montanari~\cite{Javanmard2018} also described a modified LORD procedure that achieves FDR control even under \textit{dependent} $p$-values (or equivalently, dependent test statistics). We first fix a sequence of non-negative numbers $\bm{\xi} = (\xi_i)_{i \in \mathbb{N}}$, where further conditions on $\xi_i$ are given below. This time the test levels are set as follows:\begin{equation}
\alpha_i = \xi_i W(\tilde{\tau}_i)
\label{eq:LORD_dep_potential}
\end{equation}
where $\{W(j)\}_{j \geq 0}$ represents the alpha-wealth and is defined recursively via equation~\eqref{eq:wealth_update} as before. Hence the modified LORD procedure discounts the wealth~$W$ by a factor~$\xi_i$ that depends on the number of hypotheses tested so far, while for LORD~3 the discount factor~$\gamma_{i-\tau_i}$ depends on the number of hypotheses tested since the last discovery.


For this rule, LORD controls FDR below level~$\alpha$ under a general dependency structure if $\bm{\xi} = (\xi_i)_{i \in \mathbb{N}}$ is chosen such that\begin{align}
& \sum_{j=1}^{\infty} \xi_j (1 + \log(j)) \leq \alpha/b_0 & \text{if } w_0 \leq b_0 \label{eq:LORD_dep_seq} \\ 
& \sum_{j=1}^{\infty} \xi_j (w_0 + b_0 \log(j)) \leq \alpha & \text{if } w_0 > b_0
\end{align}

The derivation in~\cite{Javanmard2018} covered the case when $w_0 \leq b_0$, and in Appendix~\ref{Asec:xi_cond} we carry out a similar analysis for when $w_0 > b_0$ to give the result above. In Section~\ref{subsec:seq_examples} we give some concrete examples of sequences of $\bm{\xi}$.

\paragraph{Upper bound on the number of hypotheses}
\hspace{6pt} If there is an upper bound $N$ on the number of hypotheses tested, then we can choose $\bm{\xi} = (\xi_i)_{i \in \{1, \ldots, N \}}$  such that \begin{align*}
& \sum_{j=1}^{N} \xi_j (1 + \log(j)) \leq \alpha/b_0 & \text{if } w_0 \leq b_0 \\
& \sum_{j=1}^{N} \xi_j (w_0 + b_0 \log(j)) \leq \alpha & \text{if } w_0 > b_0
\end{align*}

\noindent Suppose the upper bound~$N$ may need to be changed to $N'$ immediately after $n< N$ hypotheses have been tested. This requires the choice of new numbers  $\bm{\xi'} = (\xi_i')_{i \in \{n+1, \ldots, N' \}}$ such that \begin{align*}
& \sum_{j=n+1}^{N'} b_0 \xi_j' (1 + \log(j)) = \alpha/b_0 - \sum_{j=1}^{n} b_0 \xi_j (1 + \log(j)) & \text{if } w_0 \leq b_0 \\
& \sum_{j=n+1}^{N'} \xi_j' (w_0 + b_0 \log(j)) = \alpha - \sum_{j=1}^{n} \xi_j (w_0 + b_0 \log(j)) &  \text{if } w_0 > b_0
\end{align*}

\subsection{LOND}
\label{subsec:LOND}

LOND stands for Levels based On Number of Discoveries, and was proposed by Javanmard and Montanari~\cite{Javanmard2015}. Given an overall significance level~$\alpha$, we first fix a sequence of non-negative numbers~$\bm{\beta} = (\beta_i)_{i \in \mathbb{N}}$, such that~$\sum_{i=1}^{\infty} \beta_i = \alpha$. We also let~$D(n)$ denote the number of  discoveries in $\mathcal{H}(n)$, i.e.\ $D(n) = \sum_{i=1}^n R_i$.

For \textit{independent} test statistics, the values of the test levels $\alpha_i$ are chosen as follows: \begin{equation}
\alpha_i = \beta_i (D(i-1) + 1)
\label{eq:LOND_indep}
\end{equation} 

\noindent LOND can be adjusted to also control FDR under \textit{dependent} test statistics. To do so, it is modified so that the test levels are chosen as \begin{equation}
\alpha_i = \tilde{\beta}_i[D(i-1)+1)]
\label{eq:LOND_dep}
\end{equation}
where $\tilde{\beta}_i = \beta_i/(\sum_{j=1}^i \frac{1}{j})$.


Since $D(n) \geq 0$ for all $n$, it immediately follows that LOND is uniformly more powerful than the equivalent Bonferroni procedure that tests the $i$-th hypothesis at test level $\beta_i$. This guarantee does \textit{not} hold for LORD, as we shall see in the simulation studies in Section~\ref{sec:simul}.

\paragraph{Upper bound on the number of hypotheses}
\hspace{6pt} If there is an upper bound $N$ on the number of hypotheses tested, then we can choose $\bm{\beta} = (\beta_i)_{i \in \{1, \ldots, N \}}$ such that $\sum_{i=1}^N \beta_i = \alpha$. If this upper bound needs to be changed to $N'$ immediately after $n < N$ hypotheses have been tested, then we simply choose new coefficients $\bm{\beta'} = (\beta'_i)_{i \in \{n+1, \ldots, N' \}}$ such that  $\sum_{i=n+1}^{N'} \beta_i' = \alpha - \sum_{i=1}^{n} \beta_i$.

%
%

\subsection{Example sequences}
\label{subsec:seq_examples}

To use LORD or LOND, we need to specify the sequence of non-negative numbers $\bm{\gamma} = (\gamma_i)_{i \in \mathbb{N}}$, $\bm{\xi} = (\xi_i)_{i \in \mathbb{N}}$ or $\bm{\beta} = (\beta_i)_{i \in \mathbb{N}}$ to use in the procedure. The choice of these sequences will affect the power of the procedure since they directly affect the test levels~$\alpha_i$.

\paragraph{Independent test statistics}

\hspace{6pt} In~\cite{Javanmard2018}, the following sequence for $\bm{\gamma} =(\gamma_i)_{i \in \mathbb{N}}$ was proposed for LORD~2 and LORD~3: \begin{equation}
\gamma_m= C \, \frac{\log(\max(m,2))}{m e^{\sqrt{\log m}}}
\label{eq:seq_indep}
\end{equation}
with $C \approx 0.07720838$. This choice of $\bm{\gamma}$ was loosely motivated by an asymptotically optimal sequence under a mixture model (specifically, a mixture of Gaussians), where the sequence maximises a lower bound on the power of LORD. 
When there is an upper bound~$N$ on the number of hypotheses to be tested, we simply choose the value of~$C$ such that \begin{equation}
\sum_{i=1}^N \gamma_i = \alpha
\label{eq:seq_indep_bounded}
\end{equation}


When there is no upper bound, in order to make a fair comparison between LOND and LORD, we set $\beta_i = \alpha \gamma_i$ for the LOND procedure.

%
\paragraph{Dependent test statistics}

\hspace{6pt} We now present some examples of sequences $\bm{\xi}$ for LORD under a general dependency structure, since this was not considered in~\cite{Javanmard2018}. Considering the case when $w_0 \leq b_0$, the sequence $\bm{\xi} = (\xi_i)_{i \in \mathbb{N}}$ needs to satisfy equation~\eqref{eq:LORD_dep_seq}. Examples of such a sequence are \begin{align}
\xi_j & =   \frac{{C}(m)}{j^m} \label{eq:LORDseq1} \\ 
\xi_j  & = \frac{\tilde{C}(\nu)}{j \log^{\nu} (\max(j, 2))} \label{eq:LORDseq2}
\end{align}
where $m > 1$, $\nu > 2$ and $C, \tilde{C}$ are chosen so that $\sum_{j=1}^{\infty} \xi_j (1 + \log(j)) = \alpha/b_0$. To give a specific example, if we set $m = 2$ and $\nu = 3$, then $C(m) \approx 0.387224 \alpha/b_0$ and $\tilde{C}(\nu) \approx  0.139307 \alpha/b_0$.

If there is an upper bound~$N$ on the number of hypotheses to be tested, then $C$ and $\tilde{C}$ can chosen so that $\sum_{j=1}^{N} \xi_j (1 + \log(j)) = \alpha/b_0$. However, unless $N$ is small, this will not make much difference to the values of~$C$ and~$\tilde{C}$. For example, if $m = 2$, $\nu = 3$ and $N = 100$, then $C(m) \approx 0.397344 \alpha/b_0$ and $\tilde{C}(\nu) \approx  0.144134 \alpha/b_0$. 
%
Therefore, as an alternative approach with an upper bound~$N$, we can set $\xi_i = \bar{C}$, where \begin{equation}
\label{eq:LORD_bounded}
\bar{C} = \begin{cases}
\displaystyle \frac{\alpha}{b_0 \sum_{j = 1}^N (1 + \log(j))} & \text{if } w_0 \leq b_0 \\[16pt]
\displaystyle \frac{\alpha}{ \sum_{j = 1}^N (w_0 + b_0\log(j))} & \text{if } w_0 > b_0
\end{cases}
\end{equation}
For example, if $w_0 \leq b_0$ then $b_0\bar{C}/\alpha \approx 0.00215638,  1.44673  \times 10^{-4}$ and $1.08567 \times 10^{-5}$ for $N = 10^2, 10^3$ and~$10^4$ respectively.



\subsection{Example of test levels}
\label{subsec:ex_test_levels}

As an example of how the test levels for the different procedures change over time, we generate 1000 independent test statistics $Z_i \sim N(\theta_i, 1)$ where $\theta_i = \sqrt{\log 1000} \approx 2.63$ with probability 1/2 (and~$\theta_i = 0$ otherwise). We set $\alpha$ to the conventional level of 0.05, and compare the test levels for the LORD and LOND procedures designed for independent test statistics, with the detailed specifications deferred to Section~\ref{subsec:spec_proc}. As benchmark comparisons, we also calculate the test levels for the Bonferroni procedure (as specified in Section~\ref{subsec:spec_proc}), as well as the uncorrected test level of 0.05.

\begin{figure}[ht!]
\centering
\includegraphics[width = 0.7\textwidth]{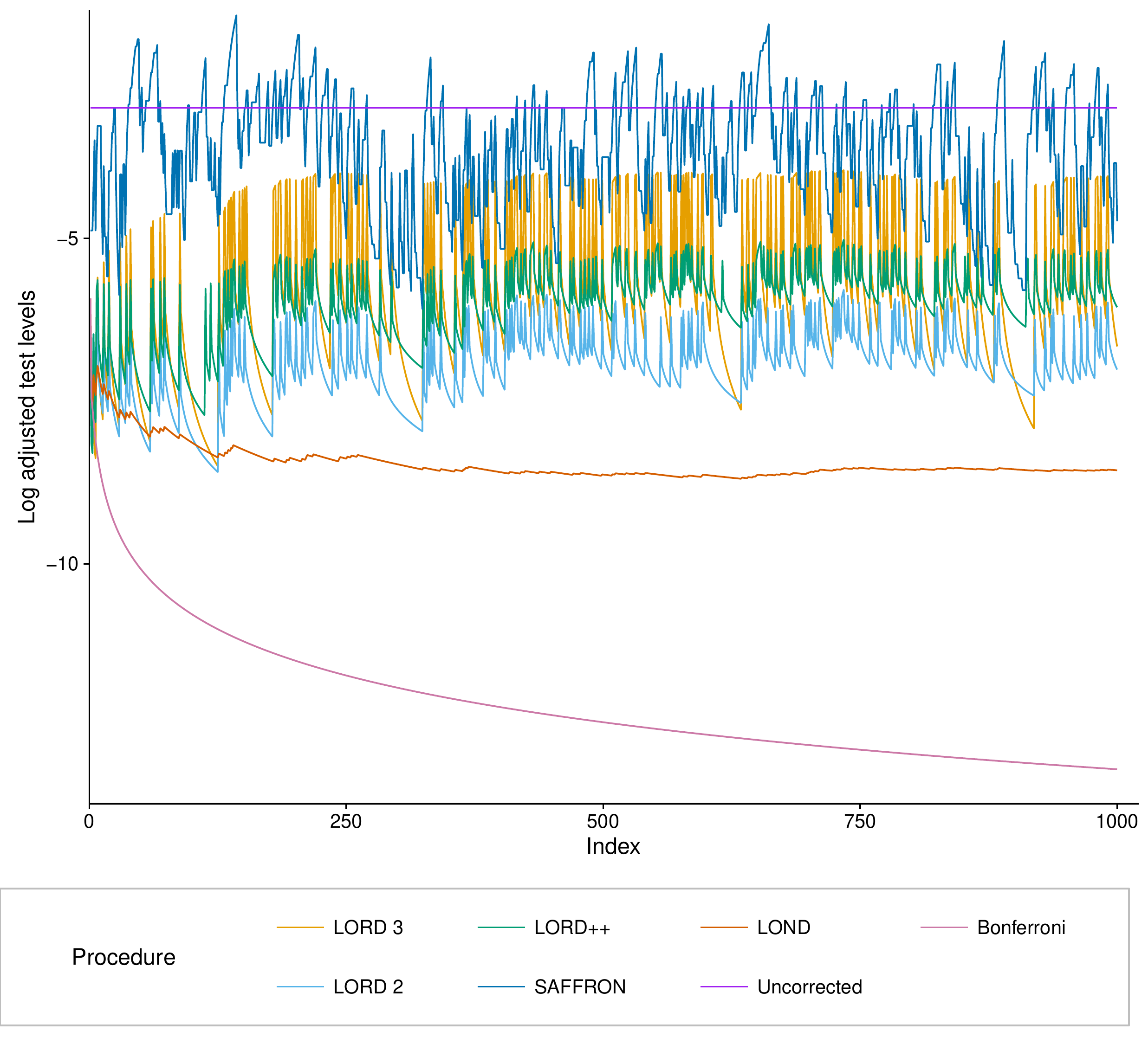} 
\caption{Log adjusted test levels for the different procedures, assuming there is no upper bound on the number of hypotheses to be tested.
\label{fig:const_unbounded_test_levels}}
\end{figure}

Figure~\ref{fig:const_unbounded_test_levels} gives the log adjusted test levels for the procedures assuming there is no upper bound on the number of hypotheses to be tested. For all procedures (except the benchmark ones), we see how the test levels increase each time a discovery is made, and otherwise decrease. In general, SAFFRON has the highest test levels, which can even be above the nominal 0.05 level. LORD~3 tends to have higher test levels than LORD++, which in turn always has test levels than LORD~2 (as discussed in Section~\ref{subsubsec:indep}). The LOND procedure has lower test levels than any of the LORD procedures after about 150 hypotheses have been tested. Finally, the Bonferroni procedure has substantially lower test levels than any other procedure once the first few hypotheses have been tested, since it can never increase its testing levels.

\begin{figure}[ht!]
\centering
\includegraphics[width = 0.7\textwidth]{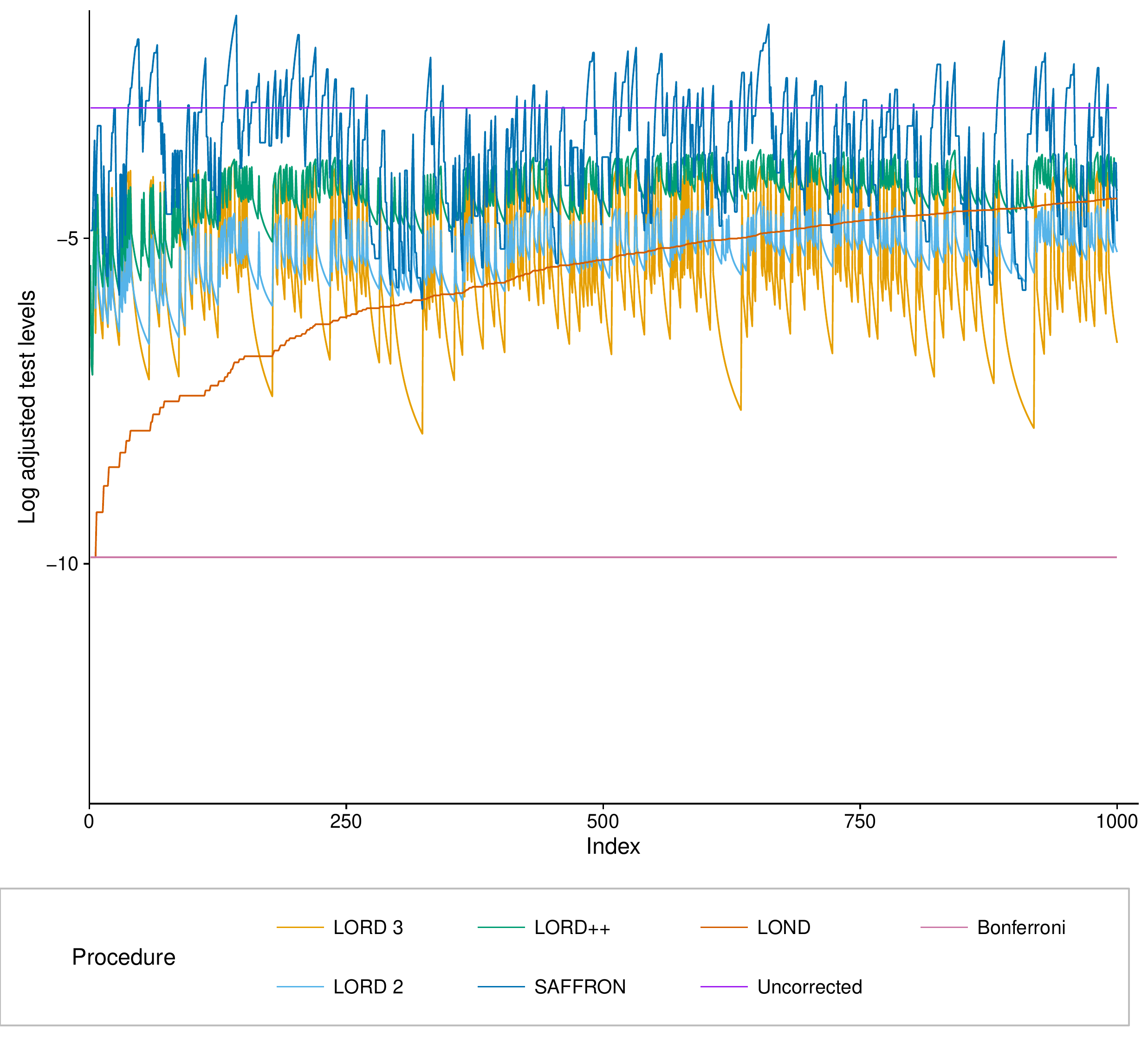} 
\caption{Log adjusted test levels for bounded procedures, assuming an upper bound of~1000 hypotheses to be tested.
\label{fig:const_bounded_test_levels}}
\end{figure}

Figure~\ref{fig:const_bounded_test_levels} gives the log adjusted test levels for the procedures assuming there is an upper bound of 1000 hypotheses to be tested. SAFFRON has virtually the same test levels as before due to the choice of $\bm{\gamma}$ (see Section~\ref{subsec:spec_proc}), and again tends to have the highest test levels. The LORD~2/3/++ procedures benefit from being able to assume an upper bound and have higher test levels than before. This time, LORD++ tends to have a higher power than both LORD~3 and LORD~2. Meanwhile, the test levels of LOND are substantially increased, and in fact become similar to those for LORD~2/3/++ as the number of hypotheses tested increase. Finally, the Bonferroni procedure always has (substantially) lower test levels than any other procedure.

\clearpage

\section{Simulation studies}
\label{sec:simul}


In this section we compare the performance of the proposed LORD and LOND procedures with the BH and Bonferroni-type procedures, in terms of the FDR and power. Our focus is on the performance with \textit{dependent} test statistics, since this was not considered for the LORD procedures in the simulation studies in~\cite{Javanmard2018, Ramdas2017, Ramdas2018}. Of particular interest is whether any FDR inflation is observed with dependent test statistics for the procedures that only provably control the FDR under independence. 

\subsection{Specification of the procedures}
\label{subsec:spec_proc}


\paragraph{Independent test statistics}

\begin{enumerate}

\item \textbf{LORD}: For LORD 2, LORD 3 and LORD++ (which for convenience, we refer to collectively from now on as LORD~2/3/++), if there is no upper bound on the number of hypotheses to be tested we use the sequence  $\bm{\gamma} =  (\gamma_i)_{i \in \mathbb{N}}$ as given in equation~\eqref{eq:seq_indep}. When there is an upper bound~$N$, we use the sequence $\bm{\gamma} = (\gamma_i)_{i \in \{1, \ldots, N \}}$ as given in equation~\eqref{eq:seq_indep_bounded}. 
For all the LORD-based methods we set the initial wealth $w_0 = \alpha/2 = 0.025$, and for LORD~2 and LORD~3 we set $b_0 = \alpha - w_0 = 0.025$.

\hspace{8pt} For SAFFRON, since it loses alpha-wealth when testing non-candidates, \cite{Ramdas2018} suggest using a more aggressive sequence (in the sense that more alpha-wealth is concentrated around the beginning of the sequence). We choose $\bm{\gamma} =  (\gamma_i)_{i \in \mathbb{N}}$ and $\bm{\gamma} = (\gamma_i)_{i \in \{1, \ldots, N \}}$ for the unbounded and bounded cases respectively, such that $\gamma_j \propto j^2$ and $\sum_{j} \gamma_j = 1$.  We follow the recommended default value of $\lambda = 0.5$ (suggested in~\cite{Ramdas2018}) .  

\item \textbf{LOND}: When there is no upper bound, we use the LOND procedure with $\bm{\beta} =  (\beta_i)_{i \in \mathbb{N}}$ set so that $\beta_i = \alpha \gamma_i$, where $\gamma_i$ is given in equation~\eqref{eq:seq_indep}, to ensure a fair comparison with LORD. Otherwise, we set $\beta_i = \alpha/N$, which we call `LOND (bounded)'.

\item \textbf{Bonferroni}: In order to be comparable to the LOND and LORD procedures described above, when there is no upper bound on the number of hypotheses, we test hypothesis $H_i$ at level $\alpha_i = \alpha \gamma_i$, where $\gamma_i$ is given in equation~\eqref{eq:seq_indep}.
%
%
When there is an upper bound $N$ on the number of hypotheses, we set $\alpha_i  = \alpha/N$ as per usual. We call this latter procedure `Bonferroni (bounded)'.

\item \textbf{BH}: We use the procedure described in Section~\ref{sec:error_rates}. In contrast to the rules described above, the BH procedure is an \textit{offline} rule (and so in our simulation studies, it is applied to all the $p$-values at once), but we use it as a `gold standard' comparison.


\item \textbf{Uncorrected}: As a comparison to all the previous rules which seek to control the FDR, we also consider the rule which does not attempt to correct for multiple testing, and simply rejects $H_i$ if $p_i < \alpha$.

\end{enumerate}


\paragraph{Dependent test statistics}

\begin{enumerate}

\item \textbf{LORD}: If there is no upper bound on the number of hypotheses to be tested, and for $w_0 \leq b_0$, we use the LORD procedure as given by equation~\eqref{eq:LORDseq2} with $\nu = 3$. When there is an upper bound~$N$, we also consider the version given in equation~\eqref{eq:LORD_bounded}, which we call `LORD (bounded)'.

\item \textbf{LOND}: When there is no upper bound, we use the LOND procedure for dependent test statistics with $\bm{\beta} =  (\beta_i)_{i \in \mathbb{N}}$ set so that $\beta_i = \alpha \gamma_i$, where $\gamma_i$ is given in equation~\eqref{eq:seq_indep}. Otherwise, we set $\beta_i = \alpha/N$, which we call `LOND (bounded)'.

\item \textbf{Bonferroni}: Since Bonferrnoi-type procedures are also valid under general dependency between the test statistics, we leave the procedure unchanged.

\item \textbf{BH}: If the dependency of the $p$-values does not satisfy a positive dependency condition (see~\cite{Benjamini2001}), the threshold for BH needs to be adjusted by replacing $\alpha$ with $\alpha/\sum_{i=1}^N \frac{1}{i}$. We call this latter variant `BH (adjusted)'.

\item \textbf{Uncorrected}: As before, we consider the uncorrected rule which rejects $H_i$ if $p_i < \alpha$.

\end{enumerate}

\subsection{Multivariate normal simulation setup}

We use a similar setup to that described in~\cite{Javanmard2015} in order to generate $N$ dependent $p$-values. Consider the hypotheses $\mathcal{H}(N) = (H_1, H_2, \ldots, H_N)$ concerning the means of normal distributions. The null hypotheses are $H_j : \theta_j = 0$. The mean parameters $\theta_j$ are set according to a mixture model: \[
\theta_j \sim \begin{cases}
0  \qquad & \text{w.p. } \quad 1 - \pi_1 \\
F_1 \qquad & \text{w.p. } \quad \pi_1
\end{cases}
\]

For each simulation run, the test statistics $Z = (Z_1, \ldots, Z_N)$ are generated according to $Z \sim N(\theta, \Sigma)$, where $\theta = (\theta_1, \ldots, \theta_N)$ and $\Sigma \in \mathbb{R}^{N \times N}$ is constructed as follows: \[
\hat{\Sigma}_{ij} = \begin{cases} 1 \qquad & \text{if } i = j \\
\rho \qquad & \text{otherwise}
\end{cases}
\] Let $\Sigma = \Lambda \hat{\Sigma} \Lambda$,  where~$\Lambda$ is a diagonal matrix with uniformly random signs on the diagonal (i.e. with probability half of being $-1$ and half of being $1$). Note that since~$\Sigma$ can have negative entries, it does not satisfy the positive dependence condition and so (in theory) the adjustment to the BH procedure as described in Section~\ref{subsec:spec_proc} is needed. 

The individual one-sided $p$-values are given by $p_j = \Phi(-Z_j)$, and two-sided $p$-values by $p_j = 2\Phi(-|Z_j|)$, where $\Phi$ is the distribution function of a standard normal variable. We have the following three choices of the non-null distribution: \begin{enumerate}

\item \textit{Gaussian}: $F_1 \sim N(0, \sigma^2)$, where $\sigma^2  = 2\log N$ (two-sided hypothesis testing).

\item \textit{Exponential}: $F_1 \sim \text{Exp}(\lambda)$, with mean $\lambda = \sqrt{2 \log N}$ (one-sided hypothesis testing).

\item \textit{Constant}: $F_1 = \sqrt{k\log N}$ (one-sided hypothesis testing), where $k = 2$ for $N \leq 100$ and $k = 1$ otherwise.

\end{enumerate}

\noindent In our simulations, we set $\alpha$ to the conventional level of $0.05$, $\rho = 0.5$ and choose the number of hypotheses $N \in \{50, 100, 1000\}$. We vary the fraction of non-null hypotheses $\pi_1$ from 0.01 to 0.09 in steps of 0.01, and then from 0.10 to 1.00 in steps of 0.05. For each value of $\pi_1$, we run $10^4$, $10^5$ and $2\times 10^5$ independent replicates for $N = 1000$, $N = 100$ and $N = 50$ respectively. \\

\subsubsection{Comparison of procedures designed for independent test statistics}
\label{subsubsec:compar_indep}

\paragraph{Unbounded procedures}

\begin{figure}[ht!]
\centering
\includegraphics[width = \textwidth]{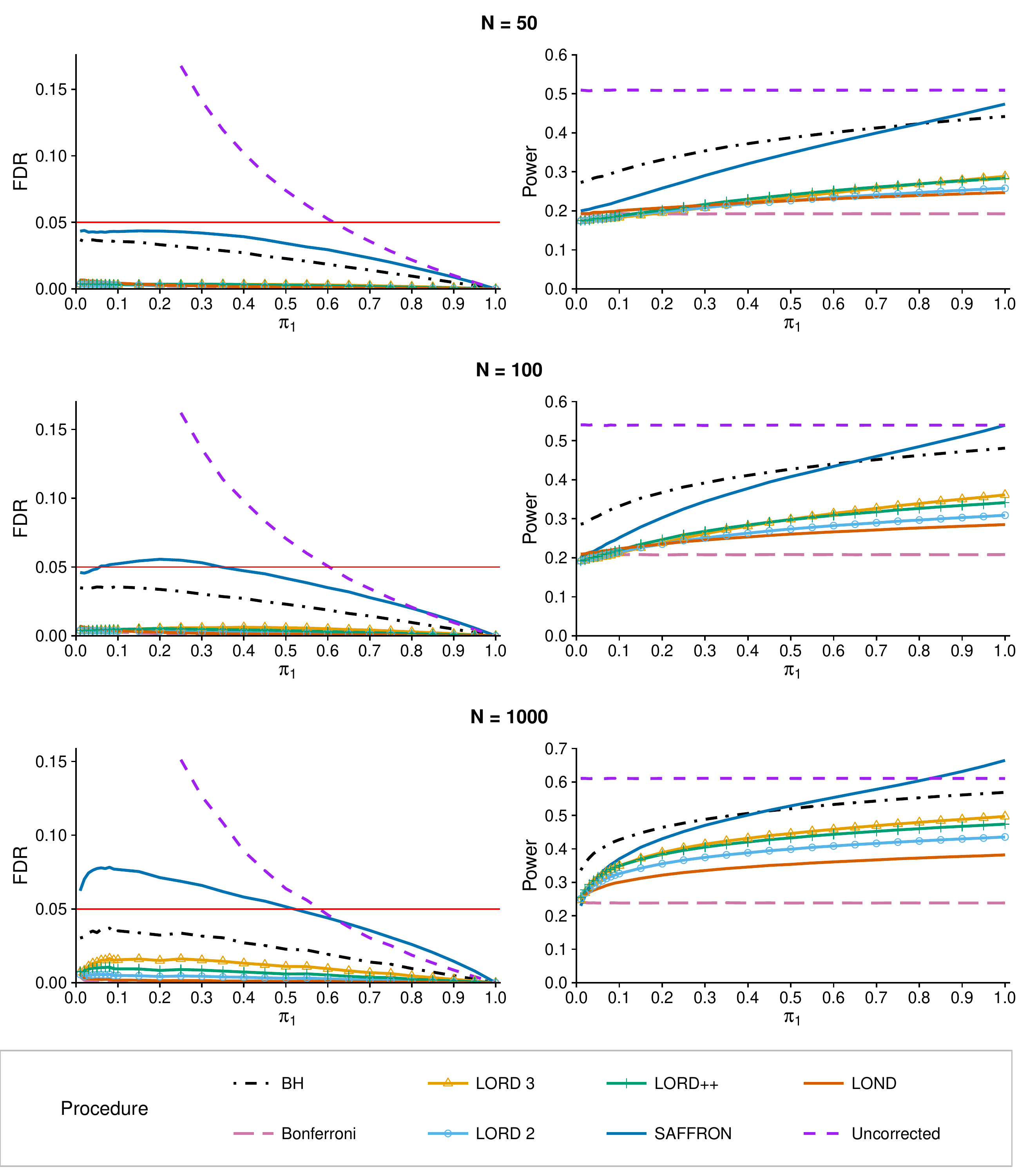} 
\caption{\textbf{Gaussian alternative (unbounded procedures)} -- FDR and power versus fraction of non-null hypotheses $\pi_1$.
\label{fig:gaussian_indep_unbounded}}
\end{figure}

\hspace{6pt} We start with comparing procedures designed for independent test statistics that are unbounded (i.e.\ no assumed upper bound on the total number of hypotheses to be tested). Figure~\ref{fig:gaussian_indep_unbounded} shows the FDR and power for the unbounded procedures under the Gaussian alternative for $N \in \{50, 100, 1000\}$.

For all values of $N$, the uncorrected test has an inflated FDR for $\pi_1 < 0.6$, with an inflation of more than 100\% for $\pi_1 < 0.35$.  This highlights the critical importance of correcting for multiple testing, particularly when there is a low proportion of non-nulls. As might be expected, for $N \in \{50, 100\}$ the uncorrected test has a higher power than any of the other procedures for all values of $\pi_1$. For $N = 1000$, the uncorrected test has the highest power except for when $\pi_1 > 0.8$, where the SAFFRON procedure has a higher power (we comment on this below). The greatest gain in power for the uncorrected tests is for small values of $\pi_1$, which corresponds to the highest inflation in the FDR.

The BH procedure controls the FDR for all values of $\pi_1$ and~$N$, even though the test statistics violate the positive dependency condition requried for BH to provably control the FDR. This demonstrates the robustness of the BH procedure under negative dependecies, which has also been observed previously in the multiple testing literature (e.g.\ see~\cite{Farcomeni2007, Kim2008, Clarke2009, Groppe2011}). In terms of power (and ignoring the uncorrected test), the BH procedure has the highest power for $\pi_1 < 0.8$, $\pi_1 < 0.6$ and $\pi_1<0.5$ when $N = 50, 100$ and $1000$ respectively, with SAFFRON having a higher power otherwise. In particular, BH has a substantially higher power than the LOND and LORD~2/3/++ procedures for all values of $\pi_1$, most noticeably when $\pi_1$ is small and $N \in \{50, 100\}$. This gain in power is to be expected given that BH is an offline procedure.

The LOND and LORD~2/3/++ procedures all control the FDR for all values of $\pi_1$ and~$N$, even though the test statistics are not independent. This shows that these procedures have some robustness under dependencies. Indeed, for $N \in \{50, 100\}$ the FDR of these different procedures are virtually indistinguishable and very close to zero. For $N = 1000$, the FDR of the procedures has the following ordering for all values of $\pi_1$: LORD 3 $>$ LORD++ $>$ LORD 2 $>$ LOND, which is the same ordering for the power of the procedures. For $N = 50$ and $N = 100$ we have the same ordering of power when $\pi_1 > 0.3$ and $\pi_1 > 0.15$ respectively.

The Bonferroni procedure is (as expected) very conservative, with a FDR indistinguishable from zero for all values of $\pi_1$ and $N$. It also has the lowest power of any of the procedures for all values of $\pi_1$ and $N$, except when $\pi_1 < 0.1$ and $\pi_1 < 0.15$ for $N = 100$ and $N = 50$ respectively. In these latter two cases, the power of the Bonferroni procedure is virtually indistinguishable from the powers of the LOND and LORD~2/3/++ procedures. However, as noted in Section~\ref{subsec:LOND}, LOND has a uniformly higher power than the Bonferroni procedure in all cases.

Finally, the SAFFRON procedure controls the FDR for all values of $\pi_1$ when $N = 50$, but has a slight inflation of the FDR above the nominal 5\% level for $\pi_1 \in (0.05, 0.35)$ when $N = 100$. When $N = 1000$, the FDR is inflated for $\pi_1 < 0.5$, with a maximum FDR of around 8\% when $\pi_1 \approx 0.05$. These results show that in contrast to the other FDR-controlling procedures, SAFFRON is not robust to departures from independence for large values of $N$ and smaller values of $\pi_1$. A likely explanation for this behaviour is that (for smaller values of $\pi_1$) SAFFRON is effectively overestimating the proportion of non-nulls due to the correlation between the test statistics, and hence the adjusted testing levels are set too high.

In terms of power, SAFFRON has the highest power out of any of the proposed online procedures for all values of $\pi_1$. As already mentioned, SAFFRON has a higher power than even BH when $\pi_1$ is large. More surprisingly, SAFFRON even has a higher power than the uncorrected test when $N = 1000$ and $\pi_1 > 0.8$, while still controlling the FDR, which is because SAFFRON procedure can have adjusted test levels above the nominal $\alpha$, as seen earlier in Section~\ref{subsec:ex_test_levels}. 

Figures~\ref{Afig:exp_indep_unbounded} and~\ref{Afig:const_indep_unbounded} in the Appendix show the results for the unbounded procedures under Exponential and Constant alternatives respectively. The results are broadly similar to those above for the Gaussian alternative, except that SAFFRON does not exhibit an inflated FDR in these settings.

\paragraph{Bounded procedures}

\begin{figure}[ht!]
\centering
\includegraphics[width = \textwidth]{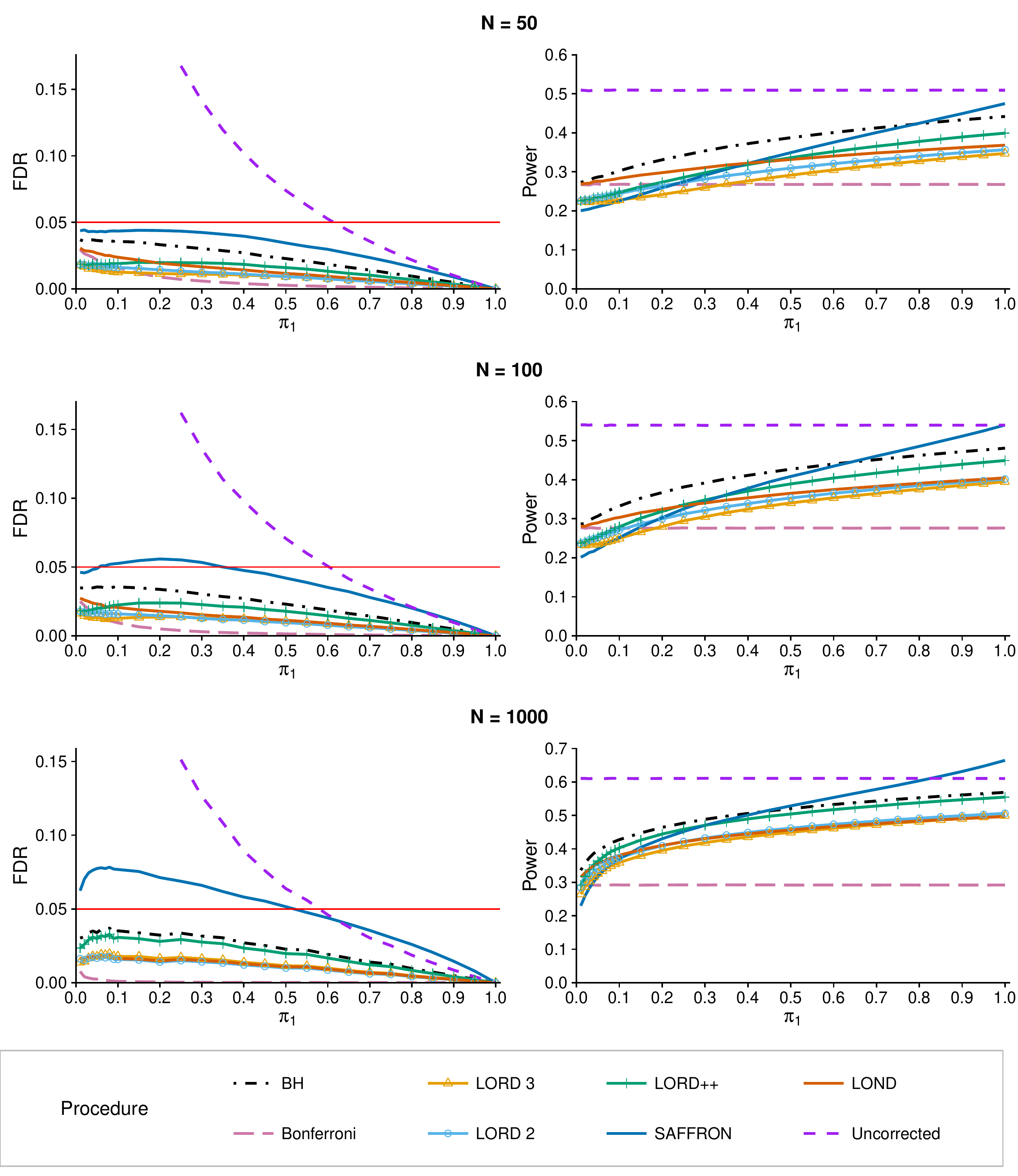} 
\caption{\textbf{Gaussian alternative (bounded procedures)} -- FDR and power versus fraction of non-null hypotheses $\pi_1$.
\label{fig:gaussian_indep_bounded}}
\end{figure}

\hspace{6pt} We now compare procedures designed for independent test statistics that are bounded (i.e.\ have an upper bound~$N$ on the total number of hypotheses to be tested). Figure~\ref{fig:gaussian_indep_bounded} shows the FDR and power for the bounded online procedures under the Gaussian alternative for $N \in \{50, 100, 1000\}$. The FDR and power for the uncorrected test and BH procedure are the same as shown in Figure~\ref{fig:gaussian_indep_unbounded}. Also, the bounded SAFFRON procedure has virtually identical FDR and power to the unbounded procedure, due to the aggressive choice of the sequence $\bm{\gamma}$ (described in Section~\ref{subsec:spec_proc}). The FDR and power of these three procedures can thus be used as reference lines when comparing the unbounded procedures with the bounded procedures in Figure~\ref{fig:gaussian_indep_unbounded} and Figure~\ref{fig:gaussian_indep_bounded} respectively.

The FDR and power of the bounded LORD 2/3/++, LOND and Bonferroni procedures is uniformly higher than the corresponding unbounded procedures (as would be expected given that the test levels are uniformly higher). However, the FDR is still controlled for all values of $\pi_1$ and~$N$, again showing the robustness of the LORD 2/3/++ and LOND procedures to departures from independence.

In terms of power, a noticeable difference from the unbounded setting is that the bounded LOND procedure now has a higher power than all other bounded procedures (apart from the uncorrected and BH procedures), including SAFFRON, for $\pi_1 < 0.4$ and $\pi_1 < 0.25$ when $N = 50$ and $N = 100$ respectively. This demonstrates the benefit of using LOND when $N$ and $\pi_1$ are relatively small. 

Another noticeable difference is that bounded LORD++ procedure has a higher power than LORD 2/3 and LOND for $\pi_1 > 0.4$ and $\pi_1 > 0.25$ when $N = 50$ and $N = 100$ respectively. When $N = 1000$, LORD++ has a higher power than LORD 2/3 and LOND for $\pi_1 > 0.05$ and in fact has a comparable power to the offline BH procedure (as well as a higher power than SAFFRON for $\pi_1 \in (0.05,  0.3)$ while still controlling the FDR for all values of $\pi_1$.

Figure~\ref{fig:gaussian_indep_bounded} also shows that for $\pi_1 > 0.4$ and $\pi_1 > 0.25$ when $N = 50$ and $N = 100$ respectively, there is the ordering LORD++ $>$ LOND $>$ LORD 2 $>$ LORD 3 in terms of power. When $N = 1000$ and  $\pi_1 > 0.05$, the powers of the LOND and LORD~2/3 procedures are very similar.

Figures~\ref{Afig:exp_indep_bounded} and~\ref{Afig:const_indep_bounded} in the Appendix show the results for the bounded procedures under Exponential and Constant alternatives respectively. The results show the same sort of patterns as discussed above for the Gaussian alternative.

\begin{figure}[ht!]
\centering
\includegraphics[width = \textwidth]{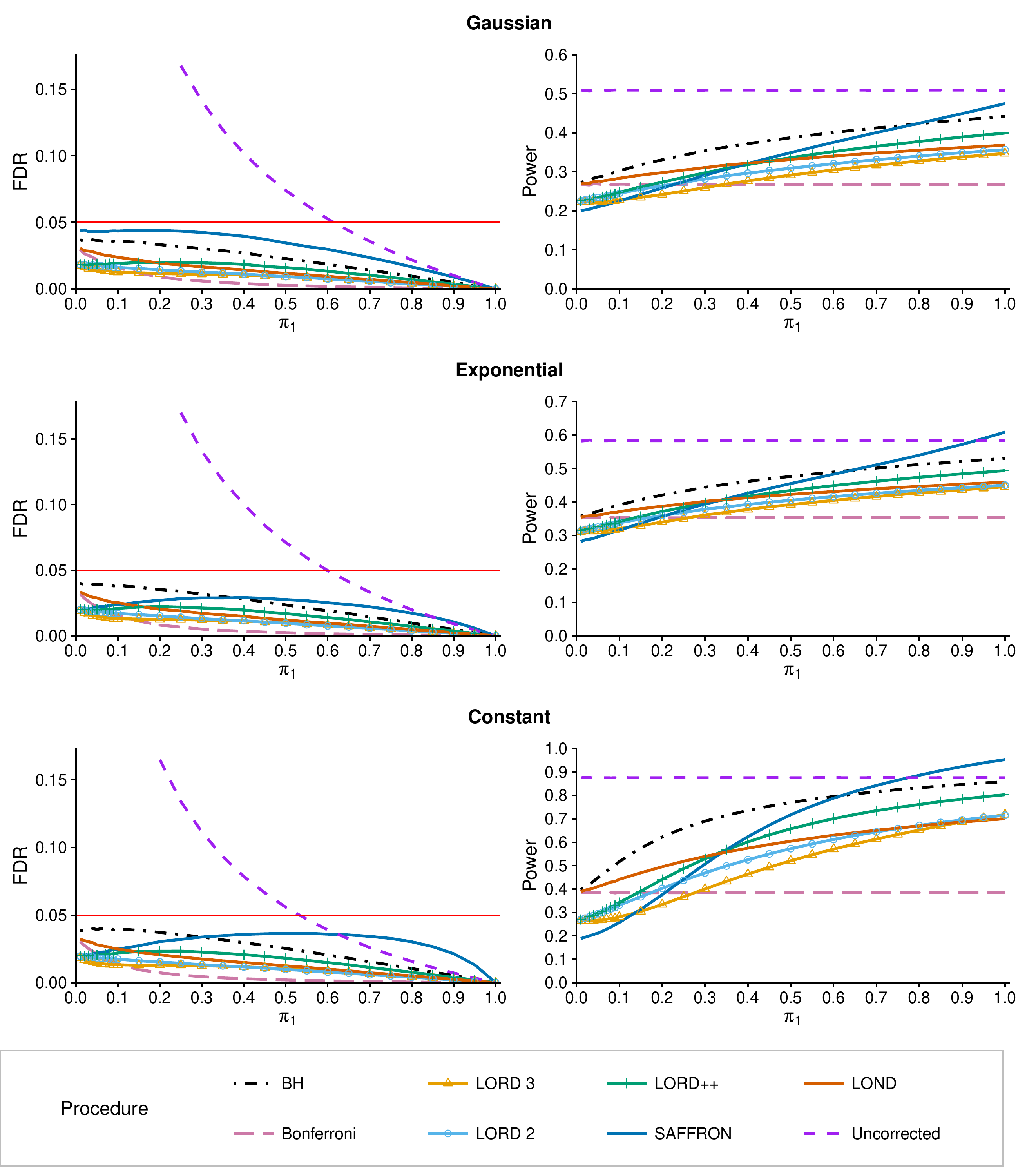} 
\caption{\textbf{Bounded procedures for $\bm{N = 50}$} -- FDR and power versus fraction of non-null hypotheses $\pi_1$.
\label{fig:N50_indep_bounded}}
\end{figure}

Focusing on the case when $N = 50$, Figure~\ref{fig:N50_indep_bounded} shows the FDR and power for the bounded procedures under Gaussian, Exponential and Constant alternatives. We again see that LOND has a higher power than all other bounded procedures (apart from the uncorrected and BH procedures) when $\pi_1<0.4$, $\pi_1 < 0.3$ and $\pi_1 < 0.35$ under Gaussian, Exponential and Constant alternatives respectively. 

Under a Constant alternative $F_1 = \sqrt{2\log(50)}$, we see that the LORD 2/3/++ and SAFFRON procedures suffer from a substantial loss of power for small values of $\pi_1$, and can have a noticeable loss of power even compared to the Bonferroni procedure. This again shows the benefit of using the LOND procedure for small values of $\pi_1$ when $N$ is also relatively small.

\subsubsection{Comparison of procedures designed for dependent test statistics}

\begin{figure}[ht!]
\centering
\includegraphics[width = \textwidth]{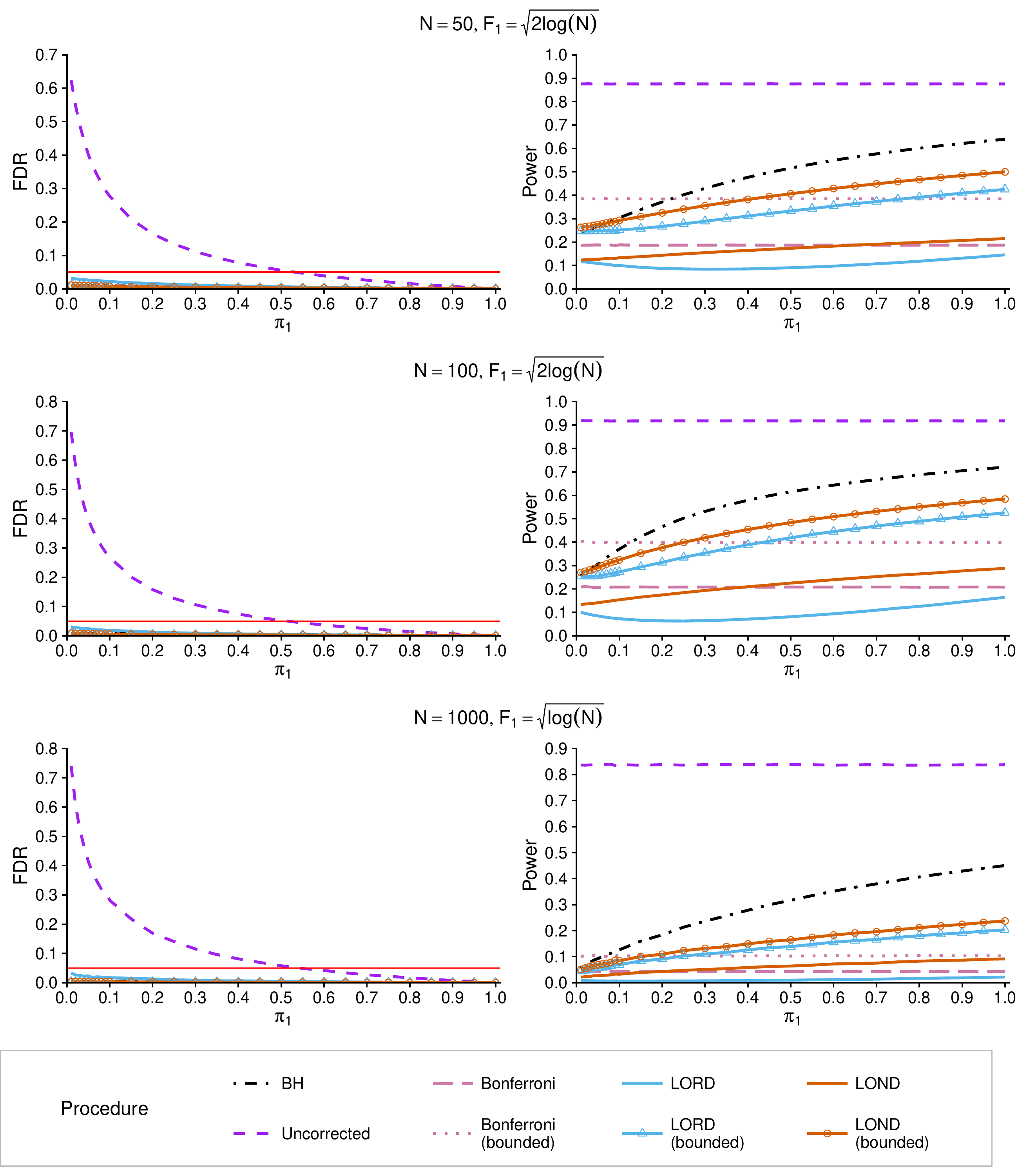} 
\caption{\textbf{Constant alternative} -- FDR and power versus fraction of non-null hypotheses $\pi_1$.
\label{fig:const_dep}}
\end{figure}

We now compare procedures designed for \textit{dependent} test statistics. Figure~\ref{fig:const_dep} shows the FDR and power for the procedures under a Constant alternative for $N \in \{50, 100, 1000\}$.

Starting with the FDR, apart from the uncorrected and unbounded LORD procedures, all other procedures have a FDR virtually indistinguishable from zero for all values of $\pi_1$. The unbounded LORD procedure has an increased FDR for $\pi_1$ close to zero, but is still always below the nominal 5\% level. In contrast, the uncorrected test has an inflated FDR for $\pi_1 < 0.5$, which increases dramatically to above 60\% as $\pi_1$ tends to zero. This again shows the importance of correcting for multiple testing when the proportion of non-nulls is low. In what follows, we do not consider the uncorrected test any further.

In terms of power, the bounded Bonferroni procedure has a higher power than any of the other procedures, including the (adjusted) BH procedure, for $\pi_1 < 0.2$ and $\pi_1 < 0.1$ when $N = 50$ and $N = 100$ respectively. Otherwise, the (adjusted) BH procedure has the highest power for $N \in \{50, 100\}$. The bounded Bonferroni procedure also has a higher power than any of the other online procedures (i.e.\ LORD and LOND) for $\pi_1 < 0.4$ and $\pi_1 < 0.25$ when $N = 50$ and $N = 100$ respectively. This suggests that for smaller values of both $\pi_1$ and $N$, there is no benefit in terms of power in using the dependent LORD and LOND procedures instead of the bounded Bonferroni procedure.

The unbounded LORD and LOND procedures in particular suffer from a very low power, with the unbounded LORD procedure having a noticeably lower power than the unbounded Bonferroni procedure for all values of $\pi_1$ and $N \in \{50, 100, 1000\}$. The unbounded Bonferroni procedure in turn has a substantially lower power than the bounded Bonferroni procedure. All this demonstrates that (regardless of which procedure used), in the dependent setting there is a substantial benefit in using a bounded instead of an unbounded procedure.

Figures~\ref{Afig:gaussian_dep} and~\ref{Afig:exp_dep} in the Appendix show the results for the procedures for dependent test statistics under Gaussian and Exponential alternatives respectively. The result are similar to those discussed above for the Constant alternative, with the LORD procedure in particular having a very low power.

\begin{figure}[ht!]
\centering
\includegraphics[width = \textwidth]{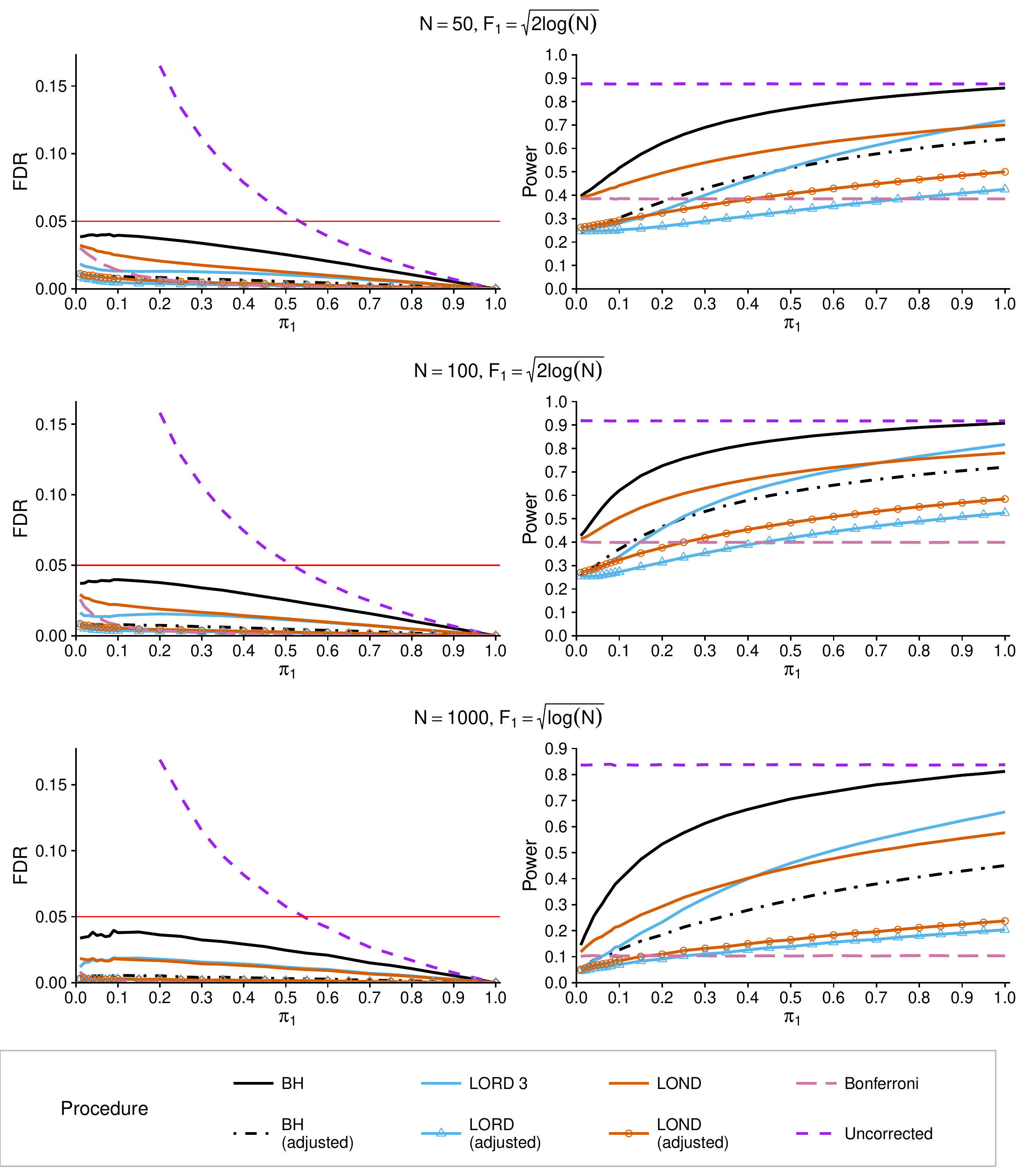} 
\caption{\textbf{Constant alternative (bounded procedures)} -- comparison of procedures designed for independent and dependent test statistics under a Constant alternative $F_1$, in terms of FDR and power versus fraction of non-null hypotheses $\pi_1$. `Adjusted' denotes the procedures designed for dependent test statistics. 
\label{fig:const_indepvsdep}}
\end{figure}

Focusing on the bounded procedures, Figure~\ref{fig:const_indepvsdep} gives a further comparison between procedures designed for independent and dependent test statistics under a Constant alternative, where `adjusted' denotes the procedures designed for dependent test statistics. We see that there is a very substantial price to pay in terms of power when using procedures designed for dependent instead of independent test statistics. Figures~\ref{Afig:gaussian_indepvsdep} and~\ref{Afig:exp_indepvsdep} in the Appendix show the same patterns for Gaussian and Exponential alternatives respectively.

%
%
%
%

\clearpage

\subsection{Platform trial scenarios}
\label{subsec:platform_trial}

We now consider a generic platform trial with normally distributed outcomes that ultimately evaluates~$K$ experimental treatments against a common control. The outcome data for patient~$i$ on the control arm is given by $Y_{i0} \sim N(\theta_0, \sigma^2)$.
The outcome data for patient~$i$ on an experimental arm~$j$ is given by $Y_{ij} \sim N(\theta_j, \sigma^2)$, where 
\[
\theta_j= \begin{cases}
\theta_0 \qquad & \text{w.p. } \quad 1 - \pi \\
\theta_0 + \sqrt{2\log K} \qquad & \text{w.p. } \quad \pi
\end{cases}
\]

At time $\tau$, there are $n_0(\tau)$ observed outcomes for patients on the control, and $n_j(\tau)$ observed outcomes for patients on experimental arm~$j$. We let $\bar{Y}_0(\tau)$ and $\bar{Y}_j(\tau)$ denote the sample means at time $\tau$ for the control arm and experimental arm~$j$, respectively, where  $\bar{Y}_j(\tau)$ is only defined when $n_j(\tau) > 0$ (i.e.\ when experimental arm~$j$ is active).

Treatment arm~$j$ is compared with the control at a fixed time~$\tau_j$ by testing the one-sided hypothesis~$H_j: \theta_j - \theta_0 \leq 0$, where it is assumed that all the control patients (including those from before treatment~$j$ is added) are used to make the comparison. We use the test statistic \[
Z_j(\tau_j) = \frac{\bar{Y}_j(\tau_j) - \bar{Y}_0(\tau_j)}{\sigma \sqrt{\frac{1}{n_0(\tau_j)}+\frac{1}{n_j(\tau_j)}}}
\] with corresponding one-sided $p$-value $p_j = \Phi(-Z_j(\tau_j))$. Due to the common control group, the test statistics and hence $p$-values for the experimental treatments will be correlated. 

For simplicity, the times $\tau_j$ (which can also be interpreted as the information fractions) are chose uniformly at random in the interval $[0.2, 1]$ for $j = 1, \ldots, K$. Given an overall sample size~$N_0$ for the control, we set $n_0(\tau_j) = \lfloor \tau_j N_0 \rfloor$. The $n_j(\tau_j)$ are chosen from a $\text{Bin}(N_{\text{target}}, 0.9)$ distribution in order to introduce some variation in the number of patients in each arm, where $N_{\text{target}}$ is the targeted sample size for each of the experimental arms in the trial.

\begin{figure}[ht!]
\centering
\includegraphics[width = \textwidth]{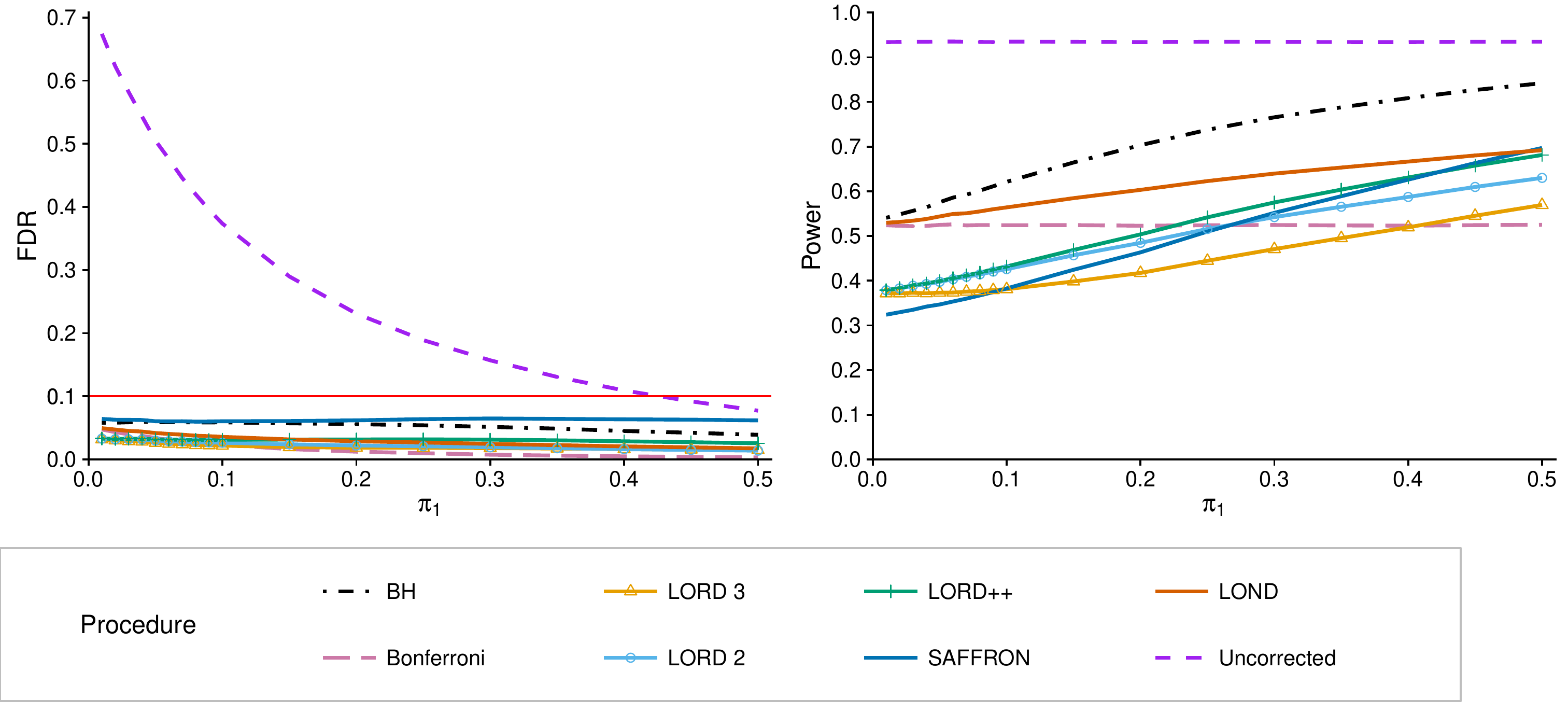} 
\caption{\textbf{Bounded procedures designed for independent test statistics} -- FDR and power versus fraction of non-null hypotheses $\pi_1$ for a platform trial that ultimately tests $K = 25$ experimental treatments. \label{fig:platform_indep_bounded} }
\end{figure}

\begin{figure}[ht!]
\centering
\includegraphics[width = \textwidth]{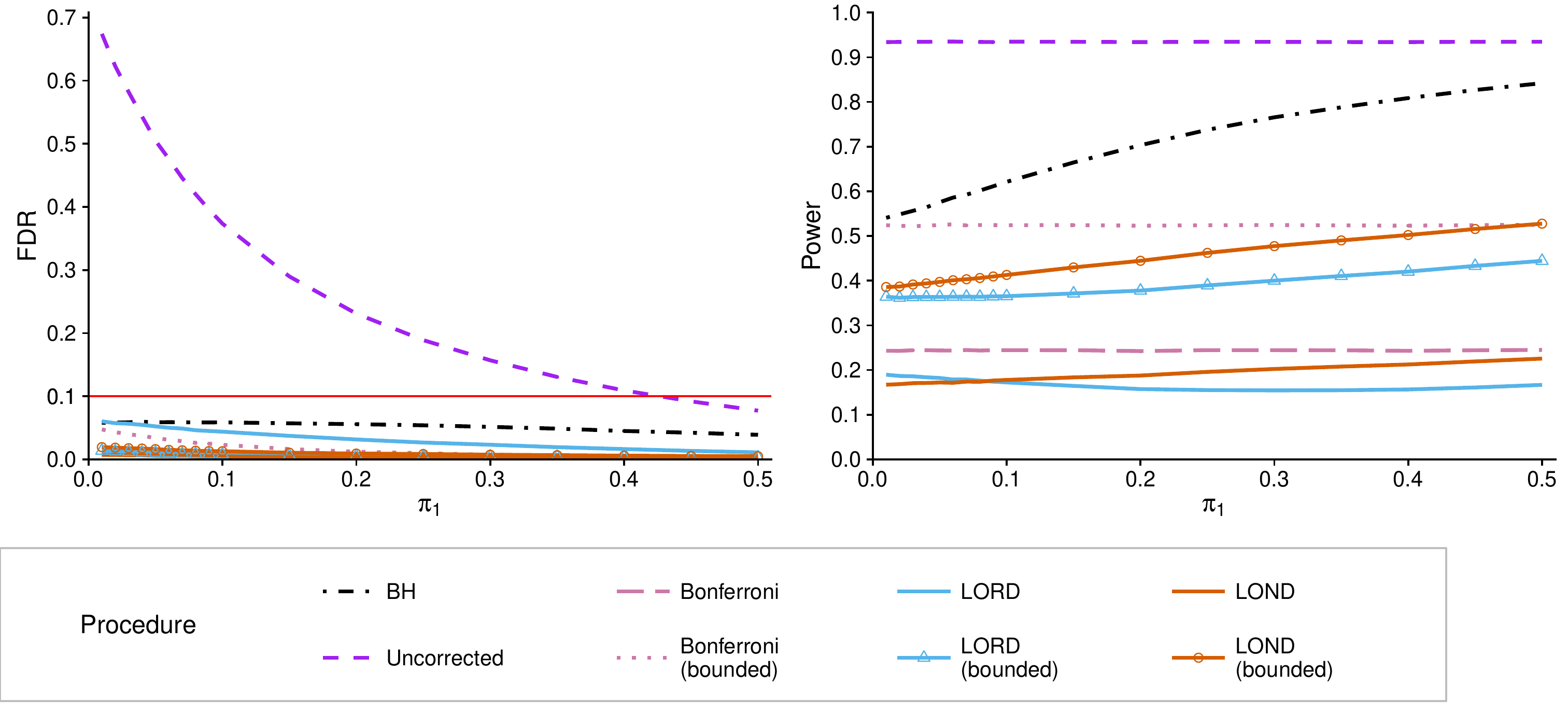} 
\caption{\textbf{Procedures designed for dependent test statistics} -- FDR and power versus fraction of non-null hypotheses $\pi_1$ for a platform trial that ultimately tests $K = 25$ experimental treatments. \label{fig:platform_dep}}
\end{figure}

In our simulation study, we set $\alpha = 0.1$,  $\sigma = 6$, $N_0 = [N_{\text{target}}\sqrt{K}]$ and $N_{\text{target}} = 70$. We set $\alpha$ higher than the usual 5\% level to reflect the possibility of using a more relaxed error rate in a platform trial (hence giving a higher power), given that treatments which `graduate' from the trial will be tested in a follow-up confirmatory study. We vary the fraction of non-null hypotheses $\pi_1$ from 0.01 to 0.09 in steps of 0.01, and then from 0.10 to 0.50 in steps of 0.05 (we only go up to 0.50 since in most disease areas, it is unlikely that more than half of new experimental treatments are truly beneficial). For each value of $\pi_1$, we run $10^5$ independent replicates. Figures~\ref{fig:platform_indep_bounded} and~\ref{fig:platform_dep} show the results for a platform trial that ultimately tests $K = 25$ experimental treatments.

Figure~\ref{fig:platform_indep_bounded} shows the FDR and power for bounded procedures designed for independent test statistics. We do not consider unbounded procedures since they suffer from a substantially lower power, as shown in Section~\ref{subsubsec:compar_indep}. All procedures (apart from the uncorrected test) control the FDR below the nominal 10\% level, with the BH and SAFFRON procedures being the least conservative. In terms of power, we see that LOND has a higher power than any of the other online procedures, and is also uniformly more powerful than the Bonferroni procedure. In turn, the Bonferroni procedure has a higher power than LORD 2/3/++ and SAFFRON when $\pi_1 < 0.25$. 

Figure~\ref{fig:platform_dep} shows the FDR and power for procedures designed for dependent test statistics. Note that since there is a positive correlation induced between the test statistics by having a common control, the BH procedure does not need to be adjusted in this setting. Again, all procedures (apart from the uncorrected) test control the FDR, with the BH and LORD procedures being the least conservative. In terms of power, the bounded and unbounded Bonferroni tests are uniformly more powerful than the corresponding LORD and LOND procedures, showing that there is no gain in terms of power in this setting by using the online procedures.

\clearpage

\section{Case studies}
\label{sec:case_study}

\subsection{Mouse phenotypic data}
\label{subsec:IMPC}


Our first case study uses high-throughput pheotypic data from the International Mouse Phenotyping Consortium (IMPC) data repository, which aims to generate and phenotypically characterize knockout mutant strains for every protein-coding gene in the mouse~\cite{Koscielny2013}. The IMPC database is an example of a growing dataset mentioned in Section~\ref{subsec:public_database}, since the family of hypotheses is constantly growing as new knockout mice lines are generated and phenotyping data is uploaded to the data repository.

We focus on the analysis of IMPC data performed by Karp et al.~\cite{Karp2017}, who looked at the influence of sex in mammalian phenotypic traits in both wildtype and mutants. All data sets, scripts and outputs for their paper can be found at \url{www.mousephenotype.org/data/sexual-dimorphism}. As part of their analysis, Karp et al.\ analysed the role of sex as a modifier of the genotype effect (for continuous traits) using a two stage pipeline. Stage~1 tested the role of genotype using a likelihood ratio test comparing models~\eqref{eq:full_model} and~\eqref{eq:null_model}. Similarly, stage~2 tested the role of sex using a likelihood ratio test comparing models~\eqref{eq:full_model} and~\eqref{eq:sex_null}. 
\begin{align}
Y & \sim \text{Genotype} + \text{Sex} + \text{Genotype}*\text{Sex} + \text{Weight} + (1|\text{Batch}) \label{eq:full_model}\\
Y & \sim \text{Genotype} + \text{Sex} + \text{Weight} + (1|\text{Batch}) \label{eq:sex_null}\\
Y & \sim \text{Sex} + \text{Weight} + (1|\text{Batch}) \label{eq:null_model}
\end{align}

The above procedure resulted in two sets of $N = 172\,328$ distinct $p$-values. Note that these $p$-values will \textit{not} in general be independent, due to positive and negative associations between different genes (caused for instance by linkage disequilibrium). In addition, multiple variables are being measured for the same gene, and these can be aspects of the same phenotype or be biologically correlated. Table~\ref{tab:IMPC_data} below shows the number of traits that had a significant genotype effect or were classed as having a significant sexual dimorphism (SD) using the unbounded procedures described in Section~\ref{subsec:spec_proc}, which were all carried out (with the exception of the uncorrected test) at a FDR level of 5\%.

\begin{table}[ht!]
	\centering
	
	\begin{tabular}{p{4cm} p{3cm} p{2cm}} 
	 & \textbf{Genotype} & \textbf{SD}\Bstrut \\ \hline
	\Tstrut \textbf{Uncorrected} & 35\,575 & 20\,887 \Bstrut \\
	\Tstrut \textbf{BH} & 12\,907 & 2\,084 \Bstrut \\
	\Tstrut \textbf{SAFFRON} & 11\,705 & 760 \Bstrut \\
	\Tstrut \textbf{LORD}$++$ & 8\,543 & 1\,193 \Bstrut \\
	\Tstrut \textbf{LORD 3} & 9\,685 & 1\,343 \Bstrut \\
	\Tstrut \textbf{LORD 2} & 8\,049 & 1\,088 \Bstrut \\
	\Tstrut \textbf{LOND} & 2\,905 & 206 \Bstrut \\
	\Tstrut \textbf{Bonferroni} & 795 &  60 \\[2pt] \hline
	\end{tabular}
	\caption{Number of rejections made by the online FDR-controlling procedures for the IMPC datasets. SD = Sexual Dimorphism}
	\label{tab:IMPC_data}
\end{table}

Looking first at the genotype effect, we have the following ordering of the procedures (in terms of the number of rejections): Uncorrected $\gg$ BH $>$ SAFFRON $>$ LORD~3 $>$ LORD++ $>$ LORD~2 $\gg$ LOND $\gg$ Bonferroni. We see that SAFFRON has almost as many (91\%) rejections as BH, but given the results of the simulation studies in Section~\ref{sec:simul}, this may be the result of having an inflated FDR. The LORD 2/3/++ procedures all have a somewhat smaller number of rejections compared to BH (of between 62--75\% of the number of rejections), but in contrast the LOND makes substantially fewer (e.g.\ only 23\% and 30\% of the number of rejections for BH and LORD++, respectively). This shows how the LORD 2/3/++ procedures gain substantial power over LOND for large~$N$ even when there are a relatively low proportion of non-nulls, as can be seen in the simulation studies in Section~\ref{sec:simul} when comparing $N \in \{50, 100\}$ with $N = 1000$.

For the SD data, we have the following ordering of the procedures: Uncorrected $\gg$ BH $>$ LORD~3 $>$ LORD++ $>$ LORD~2 $>$ SAFFRON $\gg$ LOND $\gg$ Bonferroni. This time, SAFFRON has a noticeably lower number of rejections than the LORD 2/3/++ procedures. Also, the LORD 2/3/++ procedures make a lower number of rejections compared to BH (of between 52\%--65\%). 

This difference can be explained by looking at the proportion of non-nulls for both cases. Although we do not know the proportion of traits that have a true genotype effect, we can use the uncorrected test to give an upper bound of approximately 21\%. Similarly, the uncorrected test gives an upper bound of approximately 12\% of traits  truly having a significant SD (this  bound is likely to be very generous given that the uncorrected test has 10 times as many rejections than even BH). Since we are in a scenario where there are far fewer non-nulls for SD compared to the genotype effect, we would expect the power of both SAFFRON and the LORD 2/3/++ procedures to decrease, as demonstrated in the simulation studies in Section~\ref{sec:simul}.


\subsection{Adaptive design for kidney cancer platform}
\label{subsec:kidney_platform}

Our second case study is based on a simplified version of a planned adaptive trial for kidney cancer called WIRE. In the WIRE trial, (a maximum of) 20 patients are allocated to each experimental treatment. The composite primary endpoint is tumour response to the treatment, which is binary. The null hypothesis for each treatments is that the probability of tumour response is less than or equal to 30\%. The planned trial will also include interim decision points with the possibility of early stopping for futility and efficacy based on a Bayesian statistical model, which we do consider in our case study for simplicity.

We assume that the outcome data for the control arm is given by $Y_0 \sim \text{Bin}(n_0, 0.3)$, where~$n_0$ is the number of patients allocated to the control. Denoting the total number of experimental arms by~$K$, the outcome data for experimental arm~$j$ ($j = 1, \ldots, K$) is given by $Y_j\sim \text{Bin}(n_j, p_j)$, where  $p_j = p_0 + \delta_j$ and $n_j$ is the number of patients allocated to treatment~$j$. For simplicity, we assume that treatment arm~$j$ is compared with the control by testing the one-sided hypothesis~$H_j: p_j > 0.3$ using a one-sided Fisher's exact test.

%
%

Suppose that the trial has $K = 10$ experimental arms. We set $n_0 = 32$, $n_j = 20$ ($j = 1, \ldots, 10$) and the true treatment differences $\delta = (-0.01, -0.02, 0.23, 0.52, -0.04, 0.38, -0.03, 0.22, -0.02, -0.05)$. We also set the target FDR level at 10\%.
Below are five trial realisations with these parameters: \begin{enumerate}[label=\textbf{\arabic*})]

\item $Y_0 = 19 \,$, $\; Y = (5,    5,    9,   19,    4,   15,    4,   10,    5,     6)$

\item $Y_0 = 16 \,$, $\; Y = (7,    6,   13,   14,    8,   16,    7,   11,    6,     5)$

\item $Y_0 = 13 \,$, $\; Y = (5,   13,   10,   15,   10,   15,    5,   11,    5,     3)$

\item $Y_0 = 14 \,$, $\; Y = (10,    5,    9,   17,   10,   12,   10,   15,    7,     8)$

\item $Y_0 = 14 \,$, $\; Y = (4,   10,   14,   17,    4,   16,    9,    9,    3,     3)$ 

\end{enumerate}

\noindent Table~\ref{tab:kc_trial_results} gives the FDR and power of these trial realisations for the \textit{bounded} procedures described in Section~\ref{subsec:spec_proc}. The results for scenarios~1 and~2 are typical of the trial realisations with the above parameters. In these scenarios, there are no false discoveries for any of the procedures. In scenario~1, all procedures have a the same power of 2/4, except for the uncorrected test which has a power of 3/4. In scenario~2, the Bonferroni and LORD 2/3 procedures have the lowest power of 3/4, while all the other procedures have a power of 4/4. 

\begin{table}[ht!]
\setlength{\tabcolsep}{3pt}
	\centering
	\resizebox{\textwidth}{!}{
	\begin{tabular}{l p{0cm} l  C{2cm} C{2cm} C{2.5cm} C{2.5cm} C{2.6cm} C{2.7cm} C{2cm} C{2cm}}
	 & & & \textbf{Uncorr.} & \textbf{Bonf.} & \textbf{LORD 2} & \textbf{LORD 3} & 
	\textbf{LORD$++$} & \textbf{SAFFRON} & \textbf{LOND} & \textbf{BH}\Bstrut   \\ \hline \\[-6pt]
     \textbf{1.} & & \textbf{FDR} & 0/3 & 0/2 & 0/2 & 0/2 & 0/2 & 0/2 & 0/2 & 0/2 \\ 
     & & \textbf{Power} & 3/4 & 2/4 & 2/4 & 2/4 & 2/4 & 2/4 & 2/4 & 2/4 \\ \\
     \textbf{2.} & & \textbf{FDR} & 0/4 & 0/3 & 0/3 & 0/3 & 0/4 & 0/4 & 0/4 & 0/4 \\ 
     & & \textbf{Power} & 4/4 & 3/4 & 3/4 & 3/4 & 4/4 & 4/4 & 4/4 & 4/4 \\ \\
     \textbf{3.} & & \textbf{FDR} & 2/6 & 1/4 & 2/5 & 2/6 & 2/6 & 2/6 & 2/5 & 2/6 \\ 
     & & \textbf{Power} & 4/4 & 3/4 & 3/4 & 4/4 & 4/4 & 4/4 & 3/4 & 4/4 \\ \\
     \textbf{4.} & & \textbf{FDR} & 3/7 & 0/3 & 0/2 & 2/5 & 2/5 & 2/5 & 0/3 & 3/7 \\ 
     & & \textbf{Power} & 4/4 & 3/4 & 2/4 & 3/4 & 3/4 & 3/4 & 3/4 & 4/4 \\ \\
     \textbf{5.} & & \textbf{FDR} & 2/6 & 0/3 & 0/3 & 0/3 & 1/5 & 0/3 & 0/3 & 2/6 \\ 
     & & \textbf{Power} & 4/4 & 3/4 & 3/4 & 3/4 & 4/4 & 3/4 & 3/4 & 4/4 \\ \\[-6pt]
    \hline
	\end{tabular}
}
	\caption{FDR and power of five trial realisations based on the WIRE adaptive design for kidney cancer platform. Uncorr.\ = Uncorrected, Bonf. = Bonferroni}
	\label{tab:kc_trial_results}
\end{table}

The other scenarios are more extreme, but have been chosen to illustrate some other possible (albeit unlikely) results. In scenario~3, the Bonferroni procedure has the lowest FDR (of 1/4) while still having the same power (of 3/4) as the LORD~2 and LOND procedures. All the other procedures have a power of 4/4 but an increased FDR of 2/6. In scenario~4, both the Bonferroni and LOND procedures have a FDR of 0/3 and a power of 3/4. The LORD~3/++ and SAFFRON procedures also have a power of 3/4, but have a lower FDR of 2/5. LORD~2 has a lower power of 2/4 (and a FDR of 0/2), while both the uncorrected and BH procedures have a power of 4/4 but a higher FDR of 3/7. Finally, in scenario~5 the LORD++ procedure has the lowest FDR (of 1/5) while still having the same power (of 4/4) as the uncorrected and BH procedures. All other procedures have a FDR of 0/3 and a power of 3/4.

These results illustrate the variability of the FDR and power of the different procedures when the sample size is small and a relatively low number of hypotheses are tested (we return to this issue in the discussion in Section~\ref{sec:discuss}). However, we can see that across all scenarios the uncorrected test has the highest power and highest FDR (as would be expected), while the Bonferroni procedure has the lowest FDR.


\section{Recommendations and discussion}
\label{sec:discuss}


In this paper, we have explored the performance of recent proposed procedures for online FDR control through simulation studies and two case studies, which together encompass a wide range of sample sizes~$N$, proportion of non-nulls $\pi_1$, and $p$-value distributions. Our focus has been on performance under dependent test statistics, since we would not expect test statistics to be independent in many biomedical applications. On the basis of our simulation results and case studies, we can make the following recommendations for online control of the FDR in practice:

\begin{enumerate}[label=\alph*)]



\item Across our simulation studies, there was no evidence of FDR inflation when using LORD 2/3/++ or LOND with dependent test statistics. This shows that empirically, these procedures are robust to reasonable departures from independence (recall that we set the correlation $\rho$ between the test statistics at $0.5$), and under the range of non-null distributions that we considered. This is an important result, because there is a very substantial price to pay in terms of power when switching from these procedures designed for independent test statistics to the adjusted LORD and LOND procedures designed for general dependencies. 

In contrast, with a two-sided test under a Gaussian alternative, the SAFFRON procedure had an inflated FDR for smaller values of $\pi_1$. This inflation persisted and even increased as $N$ increased from 100 to 1000. Hence, despite the high power of SAFFRON (particularly in the unbounded setting where SAFFRON dominated the other procedures), it should be used with caution if the test statistics are not independent.


\item Our simulation studies also showed the value of using the bounded versions of LORD 2/3/++ and LOND, as well as the adjusted LORD and LOND procedures. The bounded versions have a uniformly higher power than the unbounded procedures, with a very substantial gain observed for small~$N$. Hence, in applications where $N$ is likely to be relatively small (i.e.\ $N < 1000$) we recommend that if possible, an upper bound for the number of hypotheses to be tested is identified and used in the online FDR-controlling procedures (we return to this issue in the discussion below).


\item We recommend the use of the bounded LOND procedure when $N$ is small (i.e.\ $N < 100$) and the proportion of non-nulls $\pi_1$ is likely to be low. In these scenarios, LOND dominates the other procedures, as can be seen in the platform trial simulation study in Section~\ref{subsec:platform_trial}. In contrast, when $N$ is large there can be a (substantial) loss in power when using LOND compared to the other procedures, as can be seen in the IMPC case study in Section~\ref{subsec:IMPC}.


\item Across our simulation studies and case studies, we see that LORD++ has a comparable or higher power to LORD~3, and in turn LORD~3 has a higher power than LORD~2 (also, it has been shown in~\cite{Ramdas2017} that LORD++ always has a higher power than LORD~2). Hence, apart from when both $N$ and $\pi_1$ are small, we would recommend the use of the LORD++ procedure, particularly since it also provably controls the FDR for independent test statistics (whereas this was only shown empirically in~\cite{Javanmard2018} for LORD~3).


\item Focusing on procedures that provably control the FDR for dependent test statistics, we see that the bounded Bonferroni procedure does surprisingly well. In particular, for small~$N$ it actually does better than the (adjusted) LORD and LOND procedures for low values of $\pi_1$, such as in the platform trial simulation study in Section~\ref{subsec:platform_trial}. We return to this issue in the discussion below.

\end{enumerate}



\noindent As noted in recommendation b) above, a higher power can be gained by using bounded procedures. In practice, this requires some estimate of an upper bound, which may be criticised as being difficult to obtain a-priori. However, in Section~\ref{sec:online_control} we showed how to accommodate a change from an upper bound of~$N$ to $N'$ say in a straightforward manner, and hence the estimate of the upper bound can be updated over time. As well, using a bounded procedure only has substantial benefits when~$N$ is relatively small (i.e. $N < 1000$). Hence, in the public biological database setting (for example) an unbounded procedure can be used. An upper bound would be more relevant in the platform trial setting, where it is more reasonable to assume some sort of upper bound (due to e.g.\ projected funding limitations, or a projected number of drugs that can be tested each year).


Some of the recommendations above also depend on having an  estimate of the proportion of non-nulls $\pi_1$, which again may be criticised as difficult to obtain a-priori. However, by looking at historical drug success rates or the hit rates in existing public biological databases (for example), it should be possible to at least have an idea of whether $\pi_1$ is likely to be `low' or `high', which combined with whether~$N$ is `large' or `small' should be enough to decide whether to use LOND or LORD++ (or even SAFFRON, if the test statistics are independent). Another possibility may be to have an adaptive framework to online FDR control which starts off with the LOND procedure (say), and calculates an estimate of $\pi_1$ (using for example the estimate of the proportion of true nulls proposed by Storey et al.~\cite{Storey2002}). If the estimate of $\pi_1$ starts to differ substantially from the initial assumed value, then a different procedure (LORD++ say) could be used from that point onwards.


We empirically showed that the FDR was controlled when using LORD 2/3/++ and LOND even under moderate dependencies between the test statistics. Further research is required in order to find and more formally characterise the dependency structures under which FDR inflation does occur. As a related issue, there is also much scope for further development of procedures that provably control the FDR under dependent $p$-values, as the current proposals can suffer from a substantial drop in power compared to the (adjusted) BH procedure and even the (bounded) Bonferroni procedure. It would perhaps be beneficial and more natural to restrict to positive dependencies between the test statistics, firstly because this is a commonly occurring framework for clinical trials with a common control group, and secondly because the BH procedure is valid under a positive dependency condition. Indeed, we see that there is a substantial loss in power for the adjusted BH procedure compared to the usual BH procedure.


Much of our focus in this paper was on scenarios where a smaller number of hypotheses are being tested (i.e.\ when $N < 1000$). Recall that the FDR is the \textit{expected} proportion of false discoveries. Particularly when~$N$ is small and the test statistics are highly correlated, there is concern that  controlling the FDR does not prevent the FDP from varying, with the actual FDP possibly being far from its expectation~\cite{Javanmard2018}. This can be seen in some of the results in Section~\ref{subsec:kidney_platform}. Instead, the False Discovery Exceedance (FDX) may be a more meaningful metric to control, where the FDX is the probability that the FDP is greater than some threshold~$t$: \[
\text{FDX}(n) = \mathbb{P}(\text{FDP}(n) \geq t)
\] 
In \cite{Javanmard2018}, procedures are also proposed that provide online control of the FDX for independent test statistics. However, further research is needed to explore the properties of theses procedures, especially given that the FDX can be a stricter criterion than the FDR, and that FDX control is much less widely used than FDR control.


In our paper, we focused on the setting where online hypothesis testing arose due to a genuine sequential nature in the testing of the hypotheses. However, another justification is when batches of $p$-values (or even the entire set of $p$-values) are available, but the investigator chooses to process the $p$-values sequentially in a particular order. By using some sort of prior knowledge or external information, it is possible that online procedures could result in more discoveries than offline algorithms, as shown in~\cite{Javanmard2015, Javanmard2018}. We did not look at the effect of ordering the hypotheses when they arrive in batches, but this would be a useful avenue of research, particularly if tailored to biomedical applications were prior knowledge and/or independent information on the hypotheses exist. 

Additionally, we did not explore the choice of the initial wealth~$w_0$, amount earned for rejecting a hypothesis~$b_0$ or the constant $\lambda$ in the specifications of LORD~2/3/++ and SAFFRON. However, preliminary simulation results suggest that the choice of these parameters is not too crucial, particularly for large~$N$. In Section~1.3 of the Supplementary Material we carry out an analysis of the test levels for the procedures in the limiting case when every hypothesis is rejected. The analysis indicates that in scenarios where most hypotheses are expected to be rejected, $b_0$ plays a more important role and hence should be set larger (or equivalently, $w_0$ should be set smaller) for LORD~2/3/++, while for SAFFRON the constant~$\lambda$ should be set larger.


Finally, throughout the paper we have been assuming that \textit{different} hypotheses are being tested at every time. However, in many applications there is the possibility that the same hypothesis may be tested repeatedly in time. This is closer to the classical group sequential framework of hypothesis testing used in some clinical trial settings, for example. There is much scope for further research into incorporating repeated testing of the same hypotheses (potentially with the option of early stopping) within the online hypothesis testing framework.


\section*{Acknowledgements}

The authors would like to thank Natasha A. Karp for useful discussions on the mouse phenotypic data used in Section~\ref{subsec:IMPC}.


\newpage

\bibliography{FDR_control}



\begin{appendices}

\section{Conditions on $\bm{\xi}$ for the modified LORD procedure}
\label{Asec:xi_cond}

Substituting into equation~(27) in~\cite{Javanmard2018}, and letting $x_{+} = \max(0, x)$, we obtain the following inequality for the FDR when $w_0 > b_0$:
\begin{align*}
\text{FDR}(n) & \leq w_0 \xi_1 + \sum_{j=2}^n \left( b_0 \xi_j + \int_{b_0}^{w_0 + b_0 (j-1)} \frac{\xi_j}{1 + ( \frac{s-w_0}{b_0})_{+}} ds \right) \\
& = w_0 \xi_1 + \sum_{j=2}^n \left( b_0 \xi_j + \int_{b_0}^{w_0} \xi_j \, ds + \int_{w_0}^{w_0 + b_0 (j-1)} \frac{b_0 \xi_j}{s - w_0 + b_0} ds \right) \\
& = w_0 \xi_1 + \sum_{j=2}^n \xi_j (w_0 + b_0 \log(j)) \\
& = \sum_{j = 1}^n \xi_j (w_0 + b_0 \log(j))
\end{align*}

Hence when $w_0 > b_0$, LORD controls the FDR below level~$\alpha$ under a general dependency structure if $\bm{\xi} = (\xi_i)_{i \in \mathbb{N}}$ is chosen such that \begin{equation*}
\sum_{j=1}^{\infty} \xi_j (w_0 + b_0 \log(j)) \leq \alpha
\end{equation*}

\newpage

\section{Additional simulation results}

\subsection{Comparison of rules designed for independent test statistics}

\begin{figure}[ht!]
\centering
\includegraphics[width = \textwidth]{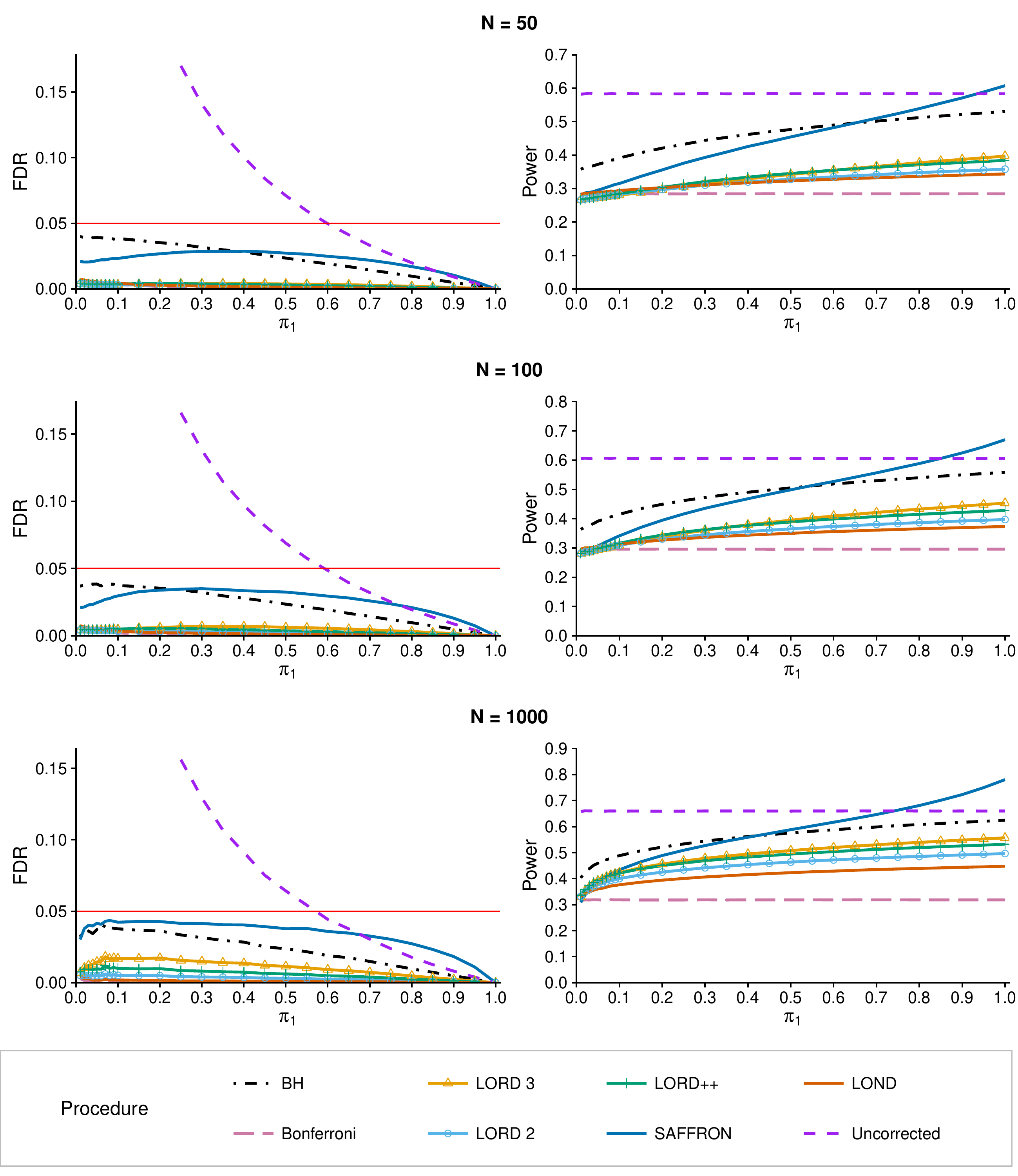} 
\caption{\textbf{Exponential alternative (unbounded procedures)} -- FDR and power versus fraction of non-null hypotheses $\pi_1$.
\label{Afig:exp_indep_unbounded}}
\end{figure}

\begin{figure}
\centering
\includegraphics[width = \textwidth]{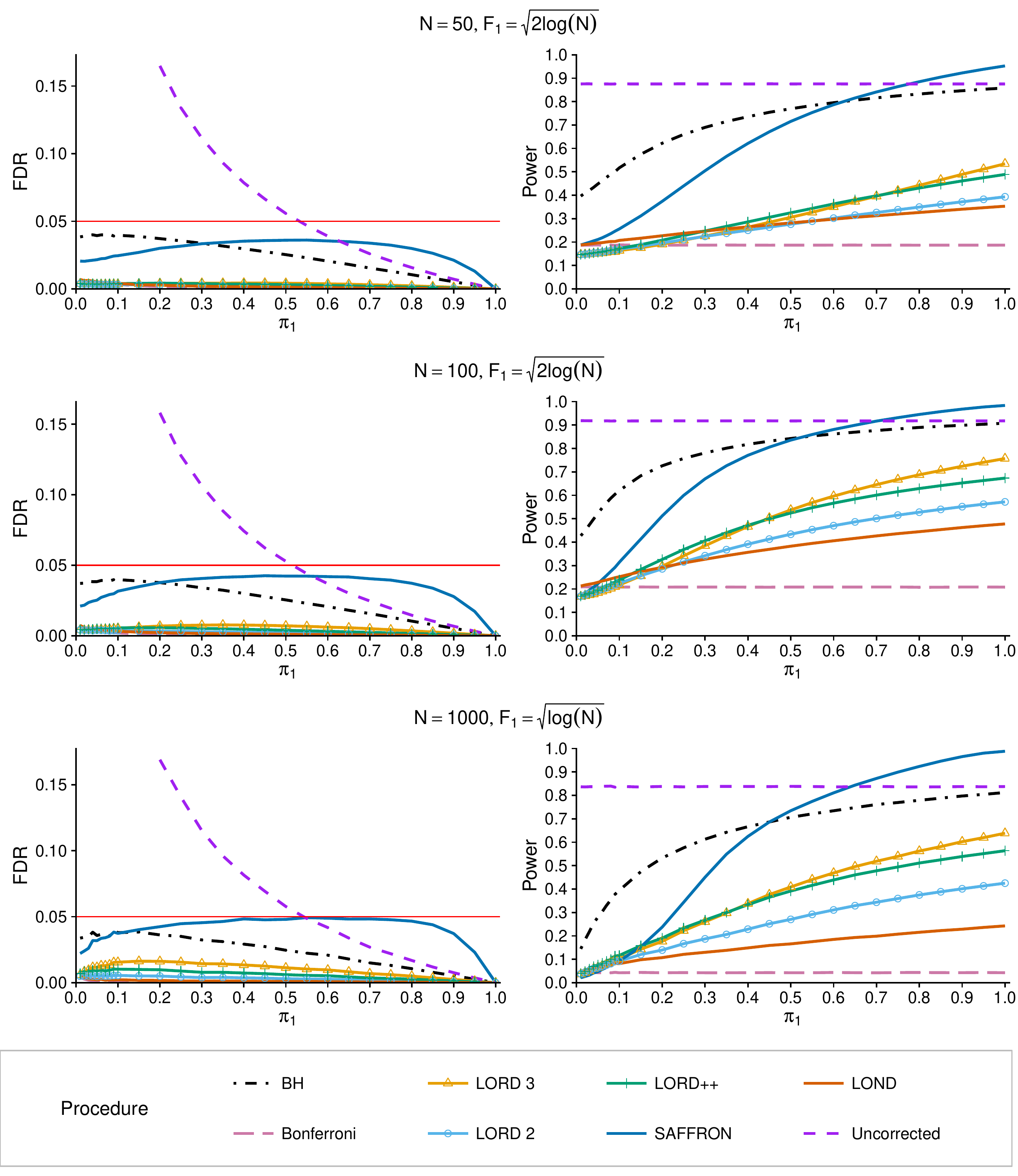} 
\caption{\textbf{Constant alternative (unbounded procedures)} -- FDR and power versus fraction of non-null hypotheses $\pi_1$.
\label{Afig:const_indep_unbounded}}
\end{figure}

\begin{figure}
\centering
\includegraphics[width = \textwidth]{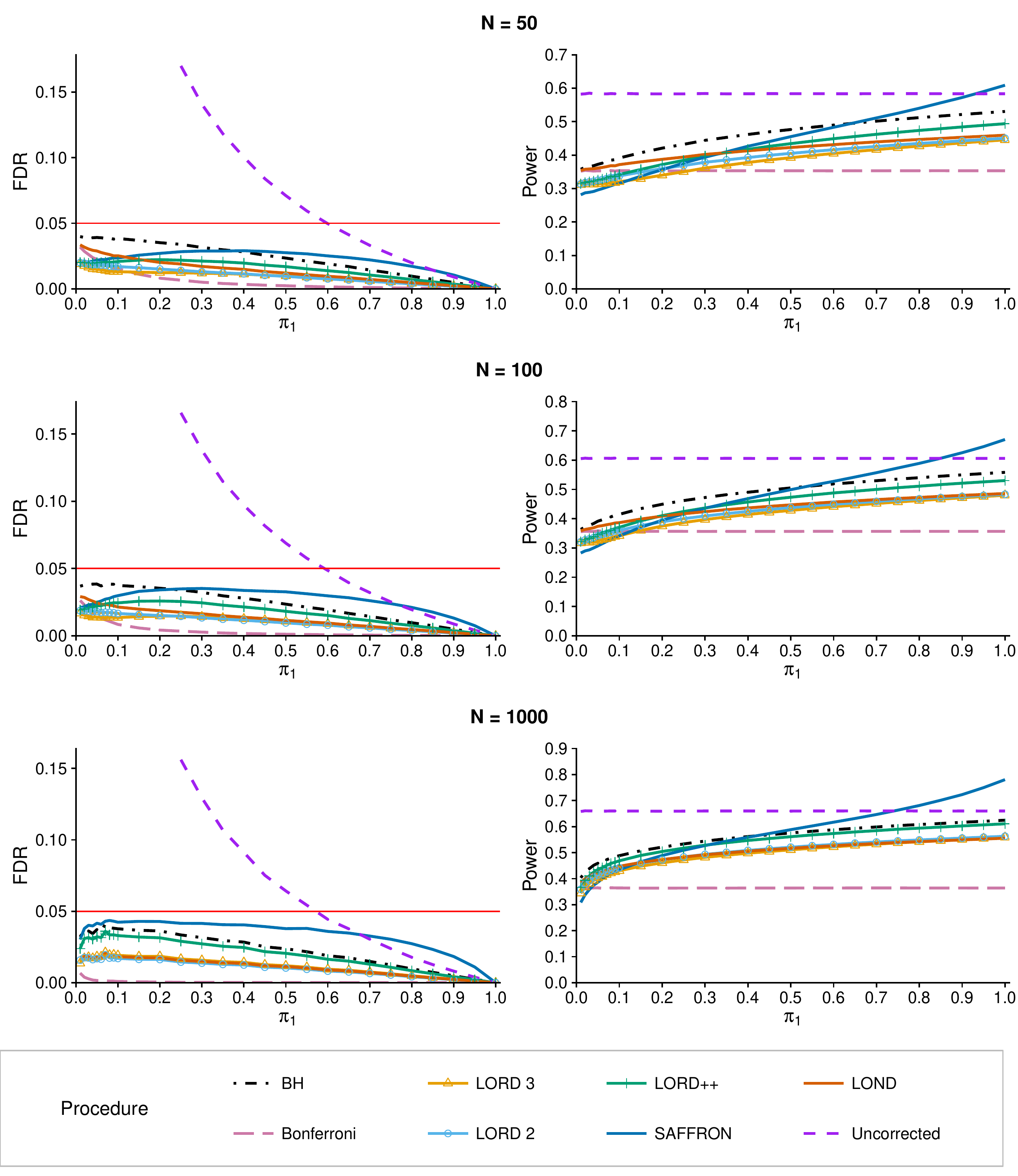} 
\caption{\textbf{Exponential alternative (bounded procedures)} -- FDR and power versus fraction of non-null hypotheses $\pi_1$.
\label{Afig:exp_indep_bounded}}
\end{figure}

\begin{figure}
\centering
\includegraphics[width = \textwidth]{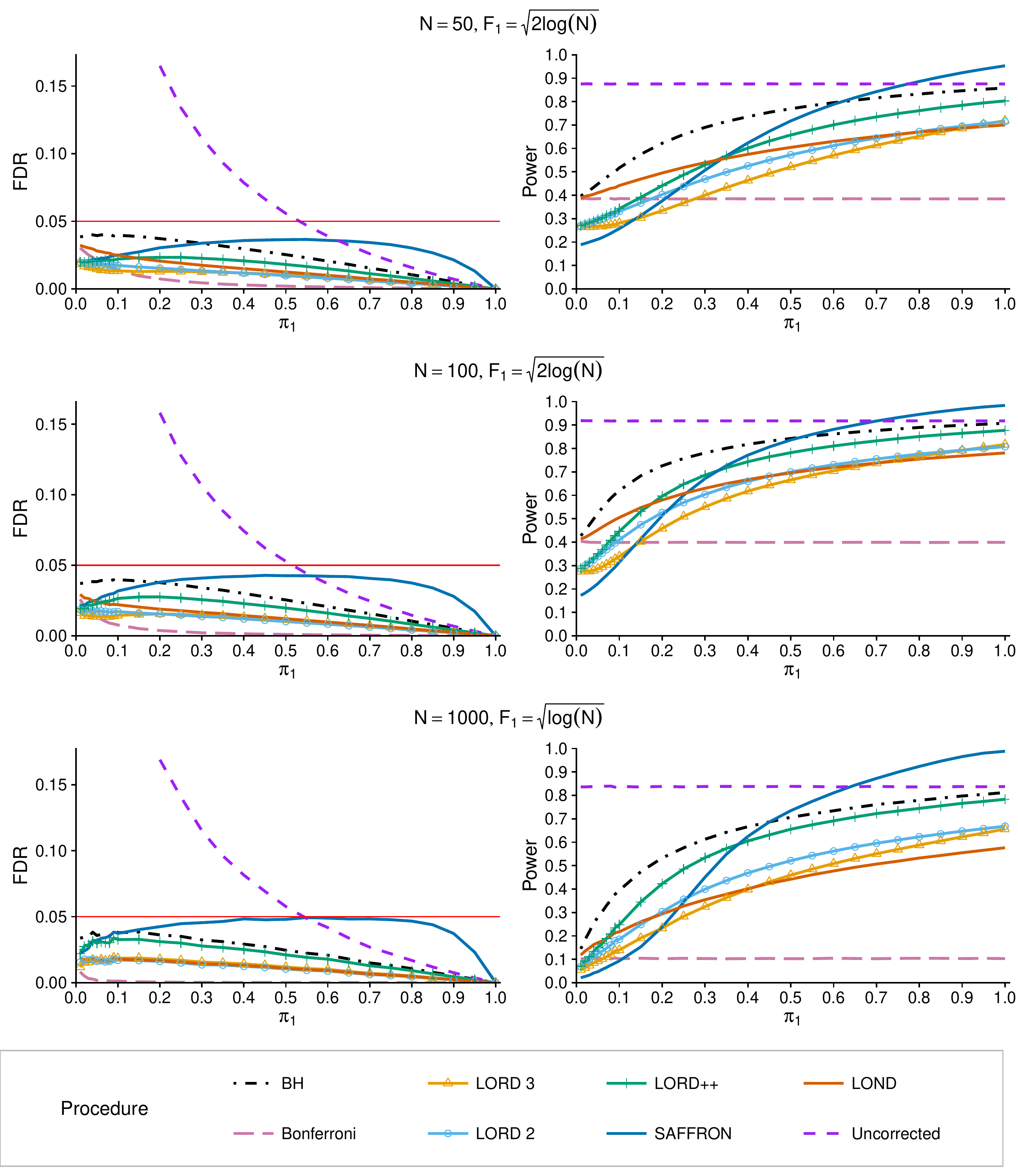} 
\caption{\textbf{Constant alternative (bounded procedures)} -- FDR and power versus fraction of non-null hypotheses $\pi_1$.
\label{Afig:const_indep_bounded}}
\end{figure}

\clearpage

\subsection{Comparison of rules designed for dependent test statistics}

\begin{figure}[ht!]
\centering
\includegraphics[width = \textwidth]{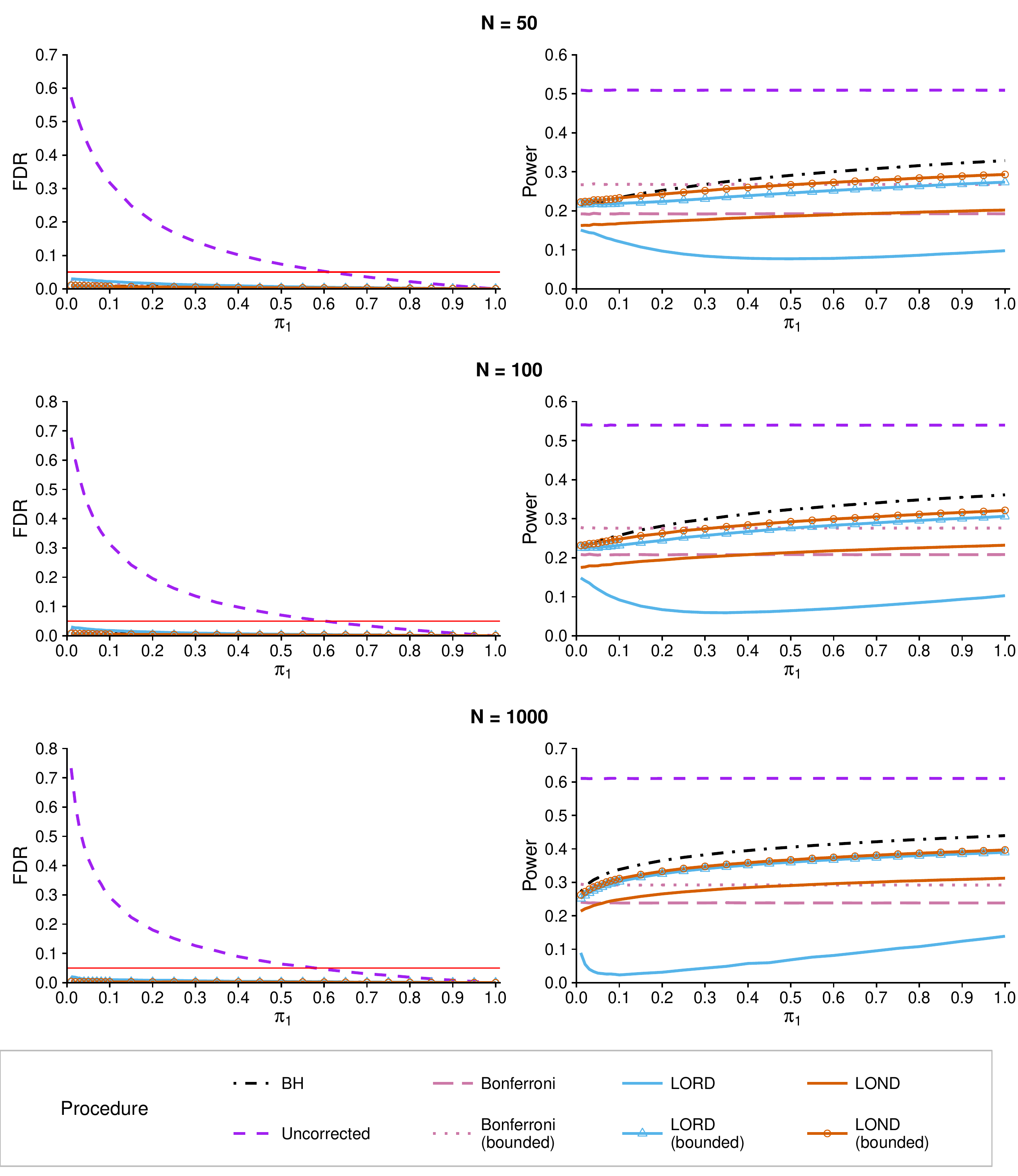} 
\caption{\textbf{Gaussian alternative} -- FDR and power versus fraction of non-null hypotheses $\pi_1$.
\label{Afig:gaussian_dep}}
\end{figure}

\begin{figure}
\centering
\includegraphics[width = \textwidth]{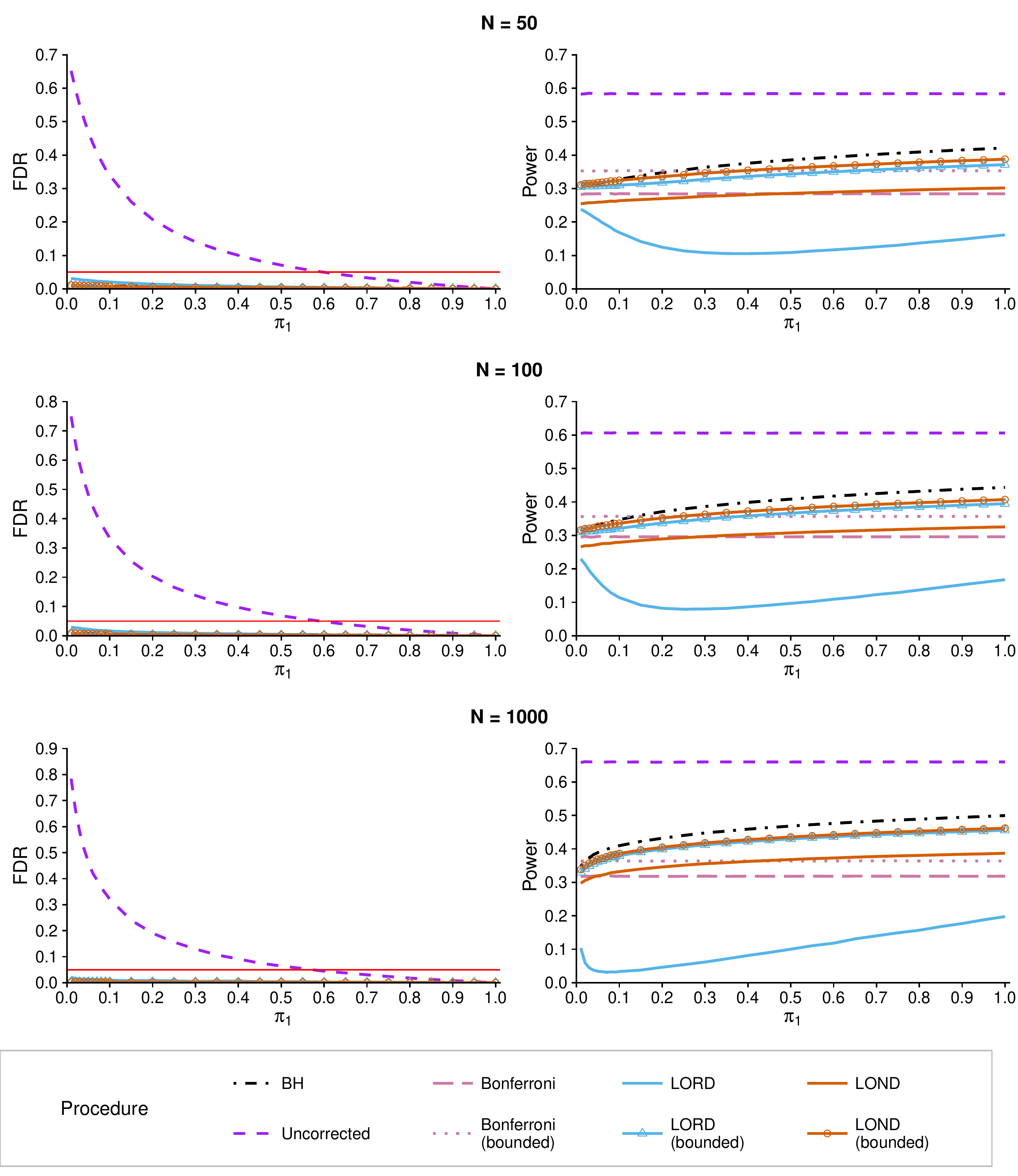} 
\caption{\textbf{Exponential alternative} -- FDR and power versus fraction of non-null hypotheses $\pi_1$.
\label{Afig:exp_dep}}
\end{figure}

\begin{figure}
\centering
\includegraphics[width = \textwidth]{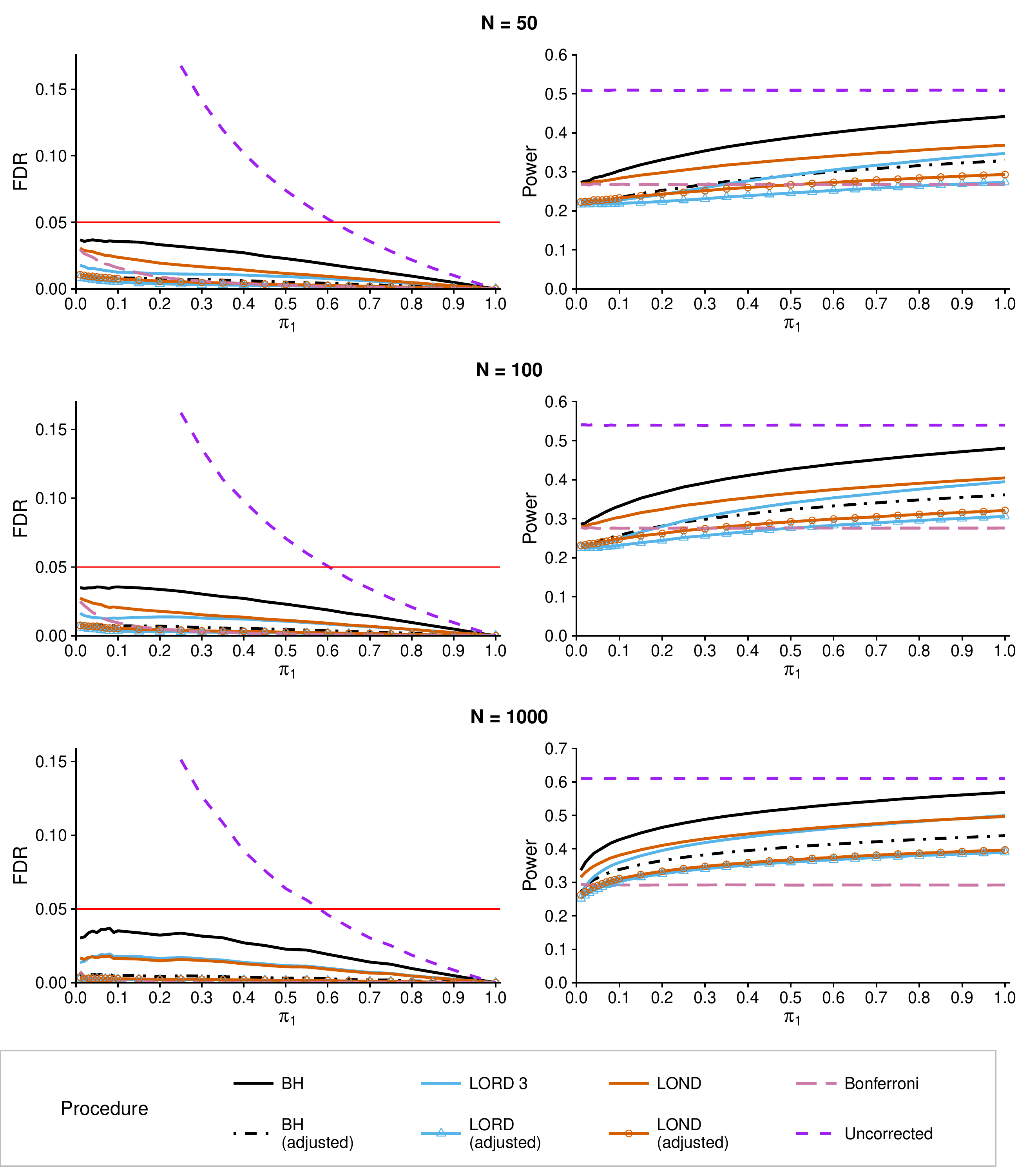} 
\caption{\textbf{Gaussian alternative  (bounded procedures)} -- comparison of rules designed for independent and dependent test statistics, in terms of FDR and power versus fraction of non-null hypotheses $\pi_1$.
\label{Afig:gaussian_indepvsdep}}
\end{figure}

\begin{figure}
\centering
\includegraphics[width = \textwidth]{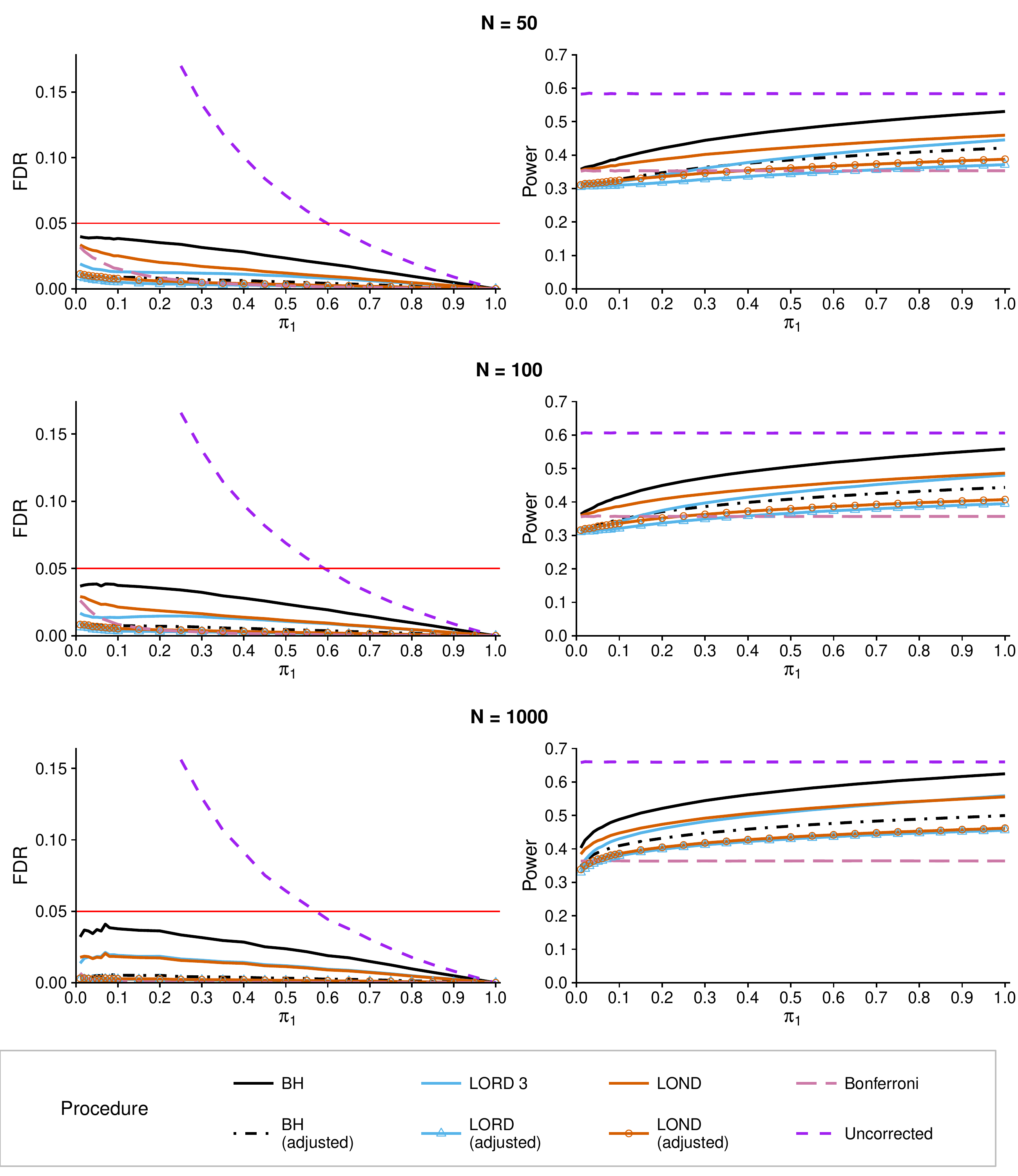} 
\caption{\textbf{Exponential alternative (bounded procedures)} -- comparison of rules designed for independent and dependent test statistics, in terms of FDR and power versus fraction of non-null hypotheses $\pi_1$.
\label{Afig:exp_indepvsdep}}
\end{figure}

\clearpage

\section{Limits of test levels}
\label{Asec:limit_test_levels}

\subsection{LORD procedures}


\paragraph{LORD 2}

If $R_i = 1$ for all $i$, then \[
\alpha_i = \gamma_i w_0 + b_0 \sum_{j=1}^{i-1} \gamma_j
\]
Hence $\alpha_i \rightarrow b_0$ as $i \rightarrow \infty$.

\paragraph{LORD 3}

If $R_i = 1$ for all $i$, then \begin{align*}
 W(i) & = (1-\gamma_1)^i w_0 + \frac{b_0}{\gamma_1}\left[ 1 - (1-\gamma_1)^i \right] \\
\implies  \alpha_i & =  \gamma_1 (1-\gamma_1)^{i-1} w_0 + b_0\left[ 1 - (1-\gamma_1)^{i-1} \right] 
\end{align*}
Hence  $\alpha_i \rightarrow b_0$ as $i \rightarrow \infty$.

%

\paragraph{LORD++}

If $R_i = 1$ for all $i$, then \[
\alpha_i = \gamma_i w_0 + (\alpha - w_0)\gamma_{i-1} + \alpha \sum_{j=1}^{i-2} \gamma_j
\]
Hence $\alpha_i \rightarrow \alpha$ as $i \rightarrow \infty$.

\subsection{SAFFRON}


If $R_i = 1$ and $C_i = 1$ for all~$j$, then
\begin{align*}
C_{j+} & = C_{j+}(i) = i-j-1 \\
\implies  \tilde{\alpha_i} & =  \gamma_1 w_0 + ((1-\lambda)\alpha - w_0) \gamma_1 + (1 - \lambda)\alpha \sum_{j=1}^{i-2}\gamma_1 \\
& =  (i-1) (1-\lambda)\alpha \gamma_1 \\
\implies \alpha_i  & = \min\{ \lambda \, , \, (i-1) (1-\lambda)\alpha \gamma_1 \}
\end{align*}
Hence $\alpha_i = \lambda$ for $i \geq 1 + \frac{\lambda}{(1-\lambda)\alpha\gamma_1}$.

%

\subsection{LORD for dependent test statistics}


If $R_i = 1$ for all~$i$, then  \begin{align*}
 W(i) & = \prod_{j=1}^i (1 - \xi_j) w_0  + \left[ 1 + \sum_{j= 2}^i \prod_{k=j}^i (1-\xi_k) \right] b_0 \\
\implies  \alpha_i & =  \xi_i \prod_{j=1}^i (1 - \xi_j) w_0  + \xi_i \left[ 1 + \sum_{j= 2}^i \prod_{k=j}^i (1-\xi_k) \right] b_0 
\end{align*}
Hence as $i$ increases $\alpha_i$ is dominated by the contribution from $ \xi_i \left[ 1 + \sum_{j= 2}^i \prod_{k=j}^i (1-\xi_k) \right] b_0 $.

\end{appendices}

\end{document}